\documentclass[11pt]{iopjournal}
\usepackage{setspace}
\usepackage{ragged2e}
\justifying
\usepackage{amsmath}

\usepackage{multicol}
\usepackage{color}
\usepackage{bm}
\usepackage{xcolor}

\usepackage{amssymb}

\usepackage{adjustbox}

\newtheorem{definition}{Definition}

\usepackage[
backend=bibtex,
sorting=none 
]{biblatex}
\begin{document}
\pagestyle{plain}
\title{Shadow models of a quantum model for cloud cover and the influence of finite sampling noise}

\affil{$^1$Deutsches Zentrum f\"ur Luft- und Raumfahrt, Institut f\"ur Physik der Atmosph\"are, Oberpfaffenhofen, Germany}
\affil{$^2$University of Bremen, Institute of Environmental Physics (IUP), Bremen, Germany}
\affil{$^*$Author to whom any correspondence should be addressed.}

\date{\today}

\author{Hedwig Keller$^{1,*}$\orcid{0009-0004-3703-6696}, Mierk Schwabe$^{1}$\orcid{0000-0001-6565-5890}, and Veronika Eyring$^{1,2}$\orcid{0000-0002-6887-4885} }

\begin{abstract}

Quantum computing is a quickly growing field that is promising various advantages compared to conventional computing. However, currently stand-alone quantum applications are scarce and hybrid (quantum-classical) computing is needed, especially in quantum machine learning (QML). Due to current limitations of quantum computing hardware and the coupling between HPC and quantum devices, integrating a trained (QML) model in classical applications is challenging. In this case, it is helpful to couple so-called shadows of the QML model instead, i.e., classical models that imitate the input-output relations of QML models such that quantum resources are only needed during the training stage. Here we consider constructive shadowing processes without an explicit training or regression stage to avoid rendering the QML model redundant, and apply them to a previously developed QML model for cloud cover \cite{Pastori.14.02.2025} to allow for an efficient coupling to a climate model.
 We compare classical interpolation methods to an approximation of the quantum Fourier model, the representation of the circuit as a partial Fourier series. The encoding strategy in \cite{Pastori.14.02.2025} allows the use of the discrete Fourier transform to efficiently reconstruct the circuits classically. Truncating the partial Fourier series further reduces the size of the shadow models. Both methods have the effect of mitigating finite sampling noise under certain conditions, which yields a motivation to use shadow models also beyond the era of limited hardware availability. Further, we compute the shadow models on the quantum system Euro-Q-Exa, based on the IQM Radiance system with superconducting qubits, where error mitigating effects can also be observed, albeit it is still difficult to distinguish them from errors connected to the calibration of the system. This work paves the way towards efficiently employing machine learning models trained on quantum computers within classical applications.

\keywords{ Quantum machine learning, Modeling, Parameterizations, Cloud cover, Classical surrogates, Error mitigation}

\end{abstract}

\section{Introduction}
Modeling is ubiquitous in science, but faces scalability issues, for instance going to high spatial and temporal resolutions. Quantum computing is emerging as an alternative or complementary approach, for example in the domain of fluid dynamics \cite{Jaksch2023}, microscopic electron transfer \cite{Gajewski2025}, or Earth system modeling \cite{Schwabe2025,Ueno.2024}. Quantum machine learning (QML) in particular has drawn significant attention in the hopes that it may provide an advantage to classical machine learning in terms of its high expressivity \cite{Schuld.2021,Benedetti.2019}, generalization \cite{Caro.2022,Caro.2023} or stability of training \cite{Klement.2026}. 

Current QML approaches are hybrid, i.e., they rely on classical computers for training. Once the QML models are trained, they are often applied in a hybrid fashion, for instance, replacing only a component of a larger classical numerical model. For example, Pastori et al.  \cite{Pastori.14.02.2025} develop a QML model for the representation of cloud cover in an Earth system model (ESM) (a so-called parametrization), with comparable performance to a classical neural network of a similar size. So far, the evaluation of the developed QML model was only done offline, i.e., without coupling it to the ESM. For QML models running natively on a quantum computer to be coupled to a classical simulation, efficient high performance computing (HPC) - quantum coupling is needed \cite{Burgholzer.2026}. However, due to the currently limited quantum hardware availability, the coupling of QML models to the classical model has to rely on a simulator of a quantum computer, which would slow down the entire simulation considerably. In this paper, we demonstrate an alternative method on the example of the above-mentioned QML cloud cover model \cite{Pastori.14.02.2025}.

 This alternative to coupling a simulated QML model was suggested in \cite{Schreiber.2023,Jerbi.2024}. The authors propose the use of so-called shadow models or classical surrogates. These are classical models, which reproduce the input-output relations of a trained QML model such that quantum hardware is only needed for training while the inference is done classically.  
 In \cite{Schreiber.2023} classical surrogates for QML models are defined, which encompass the entire hypothesis class. Popular surrogation methods are based on the quantum Fourier model, see \cite{Fontana.11.08.2022,Fontana.2025, Rudolph.17.08.2023, Sahebi.21.05.2025,Schreiber.2023, JonasLandman.}. Other publications use kernel methods \cite{Shaffer.2023,Sweke.2025}  or Gaussian processes \cite{GarciaMartin.2025} to investigate surrogation.
In \cite{Jerbi.2024} a more abstract definition beyond classical surrogates is considered which requires data generated from the quantum model. This makes a distinction regarding the dequantization of the training stage. Since this is more closely related to our problem and we are not interested in a surrogation process that allows for a dequantization of the training stage we use the term shadow model.

The risk of classical simulatibility is a general problem for the construction of shadow models. One could easily develop a data-driven model trained on data generated by the QML model. But if it can be trained also on the original training data set, the classically trained shadow model acts as a classical competitor, which both the QML and the quantum trained shadow model need to outperform to justify the quantum approach in the first place.  
This is why this work focuses on regression-free approaches, i.e., those for which a shadowing process is constructive, without the need to train the model. It instead uses the ability to evaluate the QML model on arbitrary points.

In this paper we examine two distinct approaches. First, we consider a Fourier based approach where we compute the coefficients via a fast Fourier transform (FFT), a strategy also considered in \cite{Fontana.11.08.2022,Strobl.2025,Wiedmann.05.11.2024}. To reduce the size of the shadow model, we truncate the Fourier series to the frequencies with the $N$ largest coefficients. The experiments show that for the smaller implementations of the cloud cover parametrization, approximately $10\,000$ frequencies are sufficient for a good approximation (of around $15\,000$ to $500\,000$ frequencies depending on the circuit structure). 

Secondly, we consider interpolation methods whose construction is straightforward when the basis functions and grid points are chosen appropriately. 
This approach was chosen to see how well a set function basis different from the Fourier basis might be a fit for the underlying data set. Then the interpolation might have a beneficial smoothing effect, especially once finite sampling noise from the measurements of the quantum circuit is involved.

We consider the evaluation of the circuits both without and with finite sampling noise. When comparing the QML model evaluated with finite sampling noise, compared to the shadow models constructed from the QML model evaluated with finite sampling noise, we observe that both the interpolation methods and the truncated quantum Fourier model can have an error mitigating effect. Depending on the circuit structure, the number of shots used, and the details of the shadowing procedure, the mean squared error on the test data set is lower for the shadow model compared to the QML model evaluated with shot noise on the test data. The experiments suggest that the shadowing procedures have smoothing effect on the coefficients leading to an overall mitigation of the mean squared error.

To see how well simulator-trained QML models transfer to quantum hardware, we ran some experiments on Euro-Q-Exa, a quantum computer based on IQM Radiance with superconducting qubits. We compared  the direct evaluation of the QML model trained with state vector simulation including shot noise to its shadow model. Results shows that the error-mitigating effect of the shadowing process applies to some degree to hardware noise as well as shot noise. However, it is still difficult to separate it from the underlying system error and differences due to the varying degree of the calibration of the quantum computer.

The paper is structured as follows. In Section \ref{sec:background}, we give a short overview over the relevant background, including quantum machine learning, shadow models and the cloud cover parametrization this work is based on. The methods used in the shadowing process are described in Section \ref{sec:methods}. Approximation results both simulated and on the hardware are shown in Section \ref{sec:results}. We conclude this work with a discussion in Section \ref{sec:discussion}.

\section{Background} \label{sec:background}

\subsection{A very short introduction to quantum machine learning}

Quantum machine learning is, as the name suggests, the adaptation of machine learning from classical computers to quantum computers. Trainable parameters are added to a quantum circuit, which are then optimized by hybrid classical-quantum algorithms according to a suitable loss function.
 We assume that the reader is familiar with the basics of quantum computing and its notation and forgo a more detailed introduction into QML at this point, since there is already an abundance of publications, and refer instead to \cite{Mitarai.2018,Cerezo.2021,Cerezo.2022} and for a comprehensive introduction  \cite{Schuld.2021book} and references therein.

We consider QML models of the form 
\begin{equation}
 \label{eq:exp_value}f_Q(\bm{x},\bm{\theta}) = \langle 0 \vert U^\dagger(\bm{x},\bm{\theta}) \, O \, U(\bm{x},\bm{\theta})\vert 0 \rangle
\end{equation}
which corresponds to measuring the expectation value of an observable $O$ for a trained circuit
\begin{equation}
U(\bm{x},\bm{\theta}) = W^L(\bm{\theta})S^L(\bm{x}) \cdots W^1(\bm{\theta})S^1(\bm{x})W^0(\bm{\theta})
\end{equation}
for data $\bm{x}\in \mathbb{R}^d$, trainable parameters $\bm{\theta}$, and $L$ variational layers $W^j(\cdot)$, j = \{0,\dots,L\}  and encoding layers $S^j(\cdot)$, j = \{1,  \dots L \}.

There are many different ways to encode classical data into a quantum circuit, most notably basis encoding, amplitude encoding, and Hamiltonian or angle encoding. The most common approach for QML seems to be Hamiltonian and angle encoding with a wide array of publications on data-encoding and data re-uploading available \cite{Schuld.2021,Shin.2023,Liao.2022,GilVidal.2020,Mitarai.2018, Mhiri.2024,Casas.2023}.
For Hamiltonian encoding the encoding gates are given by

\begin{equation}
S_j = e^{-i\bm{x}H_j}
\end{equation}
for a given Hermitian operator $H_j$.
A common choice for $H_j$ are the Pauli matrices (also used in \cite{Pastori.14.02.2025}), then denoted by angle encoding, which encodes the input features into single qubits as rotation angles. For $\bm{x}\in \mathbb{R}^d$ the encoding layers are then given as the unitary operations
\begin{equation}
S(\bm{x}) = \prod_{k = 1}^d e^{-i x_k \sigma_k^\alpha/2}
\end{equation}
where $\sigma_k^\alpha $, for $\alpha \in \{x,y,z\}$ is the corresponding Pauli-matrix acting on the $k$-th qubit. 
\subsection{Measurements and finite sampling noise}

To evaluate the circuit we measure the expectation value with respect to some observable $O$ as shown in Eq.~(\ref{eq:exp_value}). On simulators often an idealized setting is used, where the expectation value is computed exactly. This is no longer possible when moving to real quantum hardware where the expectation value instead has to be approximated by measuring a finite numbers of samples from the output state of the circuit, resulting in finite sampling noise, also called shot noise, where the number of shots refers to the number of samples taken. 
The noise follows a multinomial distribution, but can be modeled with a normal distribution $\mathcal{N}$ by the central limit theorem, cf. for example \cite[Methods]{Lai.2026},  such that
\begin{equation}
f^{N_s}_Q(\bm{x})= f_Q(\bm{x}) + \mathcal{N}\left(0, \frac{\sigma^2(\bm{x})}{N_s}\right)
\end{equation}
where we have taken $N_s$ samples.  The variance $\sigma^2(\bm{x})$ depends on the input data $\bm{x}$, or more specifically the state of the circuit for which we compute the expectation value.
The idealized setting, where the expectation value is computed directly, corresponds to taking an infinite number of shots. \\ 
For training QML models it is important to take this noise into account since a noisy evaluation of a circuit trained without noise results in an significant error, see Section \ref{sec:results}. However, optimizing circuits in a noisy setting has a significant impact on trainability. This  can, however, be mitigated by employing methods to regularize the variance of shot noise \cite{Kreplin.2024, Lai.2026}. The improvements that variance regularization \cite{Kreplin.2024} yield for the QML-based cloud cover parametrization are covered in \cite{Pastori.14.02.2025}.

\subsection{The quantum Fourier model} \label{sec:background:qfm}

Since we measure the expectation value to evaluate the circuit, the function $f_Q: \mathbb{R}^d \rightarrow \mathbb{R}^d$ is deterministic and real-valued, if one discounts the noise due to measurement. The function is a partial Fourier series of the form
\begin{equation} \label{eq:qfm}
f_Q(\bm{x}) = \sum_{\bm{\omega} \in \Omega} c_{\bm{\omega}} e^{-i \bm{\omega} \cdot \bm{x}}
\end{equation}
where the spectrum $\Omega$ is determined entirely be the data-encoding strategy \cite{Schuld.2021,GilVidal.2020}. In literature the sign in the exponential is often switched. Due to the symmetry of the spectrum, both notations can be used. We stick to the current notation, since it follows the common notation for the discrete Fourier transform and its inverse. 
The coefficients $\bm{c_{\omega}}\in \mathbb{C}^d$ depend on the circuit structure and the trainable parameters $\bm{\theta}$ (which we drop here for simplicity).
Since $f_Q$ is real-valued, the coefficients obey the symmetry $\bm{c_{\omega}} = \bm{c_{-\omega}}^\star$, which allows us to alternatively write

\begin{equation}
f_Q(\bm{x}) = \sum_{\omega \in \Omega^+} a_\omega \cos(\bm{x} \cdot \bm{\omega}) + b_\omega \sin(\bm{x} \cdot \bm{\omega})
\end{equation}
with  $\Omega^+ \subset \Omega$ after removing the "negative" frequencies, i.e. choosing a subset of $\Omega$, such that it 
\begin{enumerate}
\item includes the null vector $\bm{0} \in \Omega^+$,
\item is complete in the sense that $\Omega^+ \cup (-\Omega^+) = \Omega$ and
\item if $ \bm{\omega} \in \Omega^+$ and $\bm{\omega} \neq \bm{0}$, then $-\bm{\omega} \not \in \Omega^+$.
\end{enumerate}
For the one-dimensional case this simplifies to the subset with only the non-negative frequencies, i.e., if $\Omega =[-L, \dots, L]$ then we can choose $\Omega^+ = [0, \dots, L]$. The coefficients can be computed with $\bm{a_\omega} = \bm{c_\omega} + \bm{c_{-\omega}}$, $\bm{b_\omega} = -1/i(\bm{c_\omega}-\bm{c_{-\omega}})$ for $\bm{\omega} \in \Omega^+ $.

The spectrum depends on the eigenvalues of the encoding Hamiltonians, in particular it consists of the differences of all possible sums of eigenvalues of the Hamiltonian $H_j$, i.e., if we assume that $H_j$ is diagonal with
\begin{equation}
H_j = \begin{pmatrix} \lambda_{j,1} &  & \cdots & 0 \\ 0 &\lambda_{j,2} &\cdots & 0 \\ 0 & 0 & \ddots &  \\ 0 & 0 & \cdots & \lambda_{j,d_j} \end{pmatrix}
\end{equation}
then
\begin{equation}
\Omega_\ell = \left\{\Lambda_i -\Lambda_k, \bm{i,k}\in \prod_{j = 1}^L[\vert 1,d_j\vert]\right\} \text{ for } \Lambda_i = \lambda_{1,i_1} + \cdots + \lambda_{L,i_L}
\end{equation}
for the one-dimensional Fourier series, while for the multivariate case the $d$-dimensional Fourier spectrum is given by the Cartesian product over the input dimensions
\begin{equation}
\Omega = \Omega_1 \times \cdots \times \Omega_d,
\end{equation}
see e.g. \cite{Casas.2023, JonasLandman.}. In particular this has the result that for $\bm{\omega} \in \Omega$, also $-\bm{\omega} \in \Omega.$
In the simplest case, where $H_j$ are Pauli matrices, whose eigenvalues are $\pm 1/2$, this leads to an integer-valued, equidistant spectrum
\begin{equation} \label{eq:equi_spectrum}
\Omega = [-L, -(L-1), \dots, (L-1), L]^d
\end{equation}
which grows with $(2L+1)^d$ for dimension $d$ (corresponding to the number of qubits) and number of encoding layers $L$, see e.g. \cite{Casas.2023, JonasLandman.}.
To increase expressivity Hamiltonian matrices with distinct eigenvalues can be chosen, in which case the size of the frequency spectrum can grow up to $2^{2Ld}$. 
One can even add trainable parameters to their encoding layers \cite{PhysRevA.109.042421}, such that the circuit can choose the frequencies during training, or add a function, i.e. 
\begin{equation}
S_j(\bm{x}) = e^{-if_j(\bm{x})H_j}
\end{equation}
for suitable functions $f_j(\bm{x}): \mathbb{R}^d \rightarrow \mathbb{R}^m$, which allows to access a different set of basis functions than the Fourier basis \cite{Mitarai.2018,Liao.2022}. 

\subsection{Shadow models}
Our aim is the development of a shadow model on the example of the QML models for cloud cover parametrization \cite{Pastori.14.02.2025}, with the aim to eventually couple them to the climate model ICON-XPP \cite{Mueller2025b}, circumventing problems caused by limited hardware availability.

In contrast to classical surrogates \cite{Schreiber.2023}, we use the term shadow models since in the definition of  \textit{shadow models} in \cite{Jerbi.2024} an explicit requirement of data generated by a quantum model, termed as quantum advice, is made, rather than aiming at a surrogation of the training stage as well.

\begin{definition}[Shadow models and shadowfiability, cf. \parencite{Jerbi.2024}]\label{def:shadow} \rm
An $n$-qubit quantum model $f_{\bm{\theta}}$ is \textit{shadowfiable} if there exists a classical model, the \textit{shadow model} $\widetilde{f}_{\bm{\theta}}$, such that with probability $1-\delta$ over the generation of the quantum advice $ \omega(\bm{\theta})$, it holds that
\begin{equation}
\max_{\bm{x}\in \mathbb{R}^d} \vert f_{\bm{\theta }}(\bm{x}) - \widetilde{f}_{\bm{\theta }}(\bm{x}) \vert \le \varepsilon
\end{equation}
for which only $m,M \in \mathcal{O}(\mathrm{poly}(n,1/\varepsilon,1/\delta))$ qubits and circuits, respectively,  were used in the generation of the quantum advice. 
The shadow model $\widetilde{f}_{\bm{\theta}}$ is defined as
\begin{equation}
    \widetilde{f}_{\bm{\theta}}(\bm{x}) = \mathcal{A}(\bm{x},\omega(\bm{\theta}))
\end{equation}
where $\mathcal{A}$ is a classical $\mathcal{O}(\mathrm{poly}(M,m,d))$  time algorithm which processes the quantum advice $\omega(\bm{\theta})$ for the input $\bm{x} \in \mathbb{R}^d$.
\end{definition}

The work in \cite{Jerbi.2024} gives some theoretical justification and limits of shadow models. In particular it is shown that under widely assumed complexity assumptions there exist shadow models that have advantages over fully classical models, but also that not all quantum models are shadowfiable. Also, an example is given that not all shadowfiable models can be represented by the Fourier-based classical surrogate.

Considering that the models in \cite{Pastori.14.02.2025} use a simple encoding scheme such that we know a priori the frequency spectrum, the quantum Fourier model  yields an obvious choice for a shadow model. \\
One way to compute the coefficients of the QFM \eqref{eq:qfm}, assuming the spectrum $\Omega$ is known, as used in several publications \cite{Schreiber.2023, JonasLandman., Hernicht.08.08.2025}, is by choosing a set of points $X$ and computing
\begin{equation}\label{eq:least_square}
\bm{c}^\star = \mathrm{argmin}_{\bm{c} }\Vert A\bm{c} -\bm{Y} \Vert^2
\end{equation}
with $A_{j,\bm{\omega}} = e^{-i\bm{\omega} \cdot \bm{x}_j}$ and $Y = [f_Q(\bm{x}_j)]_j$ for $\bm{x}_j \in X$. In \cite{Schreiber.2023}, the points are chosen as an equally spaced grid on $[0,2\pi)$, but could also be chosen from the training data \cite{Hernicht.08.08.2025}, or simply at random.
There are several caveats to this approach.  
Firstly, one could also replace $X$ and $Y$ by the training data and get a purely classical model, which might make the quantum model redundant and perhaps outperform it \cite{Schreiber.2023}. 
Secondly, the matrix $A$ depends on the size of the spectrum and therefore grows exponentially and is non-sparse.  To counteract this, several sampling strategies with random Fourier features (RFFs) were suggested in \cite{JonasLandman.}. 
In \cite{RyanSweke.,Sweke.2025} necessary and sufficient conditions for efficient classical simulatability with RFFs were proposed. They show that one necessary condition for an efficient shadowing of an already trained circuit with RFFs is a sufficient concentration of the function in the Fourier spectrum. In particular, this means that a suitable distribution for the coefficients over the spectrum exists, which sufficiently concentrates, and can be identified and efficiently sampled.
While we have the advantage of already having a trained circuit, and not needing to dequantize the training stage as well, as is considered \cite{Schreiber.2023,Rudolph.17.08.2023}, finding how the function concentrates is difficult to achieve without evaluating the coefficients of the Fourier series on the entire spectrum.

If the circuit structure is known, the coefficients can also be computed explicitly. Some publications make use of this by choosing a strict circuit structure \cite{Rudolph.17.08.2023} albeit at the danger of classical simulatibility. However in general, this takes exponential effort with respect to the number of encoding layers \parencite{Nemkov.2023,Wiedmann.05.11.2024}.

The high expressibility is one of the strengths of quantum computing, and is desirable in general when it comes to regression. But it can cause significant problems for trainability such as barren plateaus \cite{Cerezo.2025, Cerezo.2021BP, Larocca.2025,Okumura.2023,Holmes.2022}. 
 While there are ways to avoid barren plateaus, their absence might imply classical simulatibility \cite{Cerezo.2025}. This is a general problem that quantum machine learning faces \cite{GilFuster.2024} and similarly in the development of a shadowing process different agendas need to be balanced.
For one, it needs to be efficient in its computation, while still reflecting the high expressivity of the quantum circuit. As we have seen in the derivation of the quantum Fourier model, the space of accessible functions can grow exponentially with respect to the number of input features and encoding layers. Therefore, at some point for an efficient shadowing process, it becomes necessary to reduce the function space, while possible advantages that one might observe during training are not lost.

\subsection{The cloud cover parametrization}

We shortly summarize here the most important aspects of the QML-based cloud cover parametrization of \cite{Pastori.14.02.2025}. For detailed information we refer to \cite{Pastori.14.02.2025}.
For our purposes the relevant aspects are that the models describe a cloud cover parametrization $f_{clc}: \mathbb{R}^d \rightarrow [0,1] \subset \mathbb{R}$, where the output describes the fraction of a given climate model cell covered in clouds. The parametrization consists of three parts. First, the input features are classically preprocessed by $f_{pre}$ such that they are more uniformly distributed and (nearly) contained in $[0,\pi]$. This is followed by the quantum circuit with two different ansatzes denoted by $f_{Q,ZZXY}$ and $f_{Q,XYZ}$ which map the pre-processed data from $\mathbb{R}^d$ to $\mathbb{R}^d$, and lastly a classical post-processing $ f_{post, \cdot}$, consisting of an affine linear mapping with trainable weights and biases, which maps the output to the interval $[0,1]$ and contains an inversion of the input transform of the cloud cover that is used in the training. The dependence of $f_{post,\cdot}$ on the circuit ansatz is specified, since it includes the trained weights of the affine linear mapping.
The original quantum circuit considered eight input features, but a feature analysis showed that six input features are enough to predict the cloud cover. Unless specified otherwise, we consider the problem for $d = 6$. \\
In summary, we consider two different trained QML models
\begin{equation}\label{eq:qml_clc_arch}
\begin{aligned}
f_{ZZXY}  &= f_{post,ZZXY} \circ f_{Q,ZZXY} \circ f_{pre} : \mathbb{R}^6 \mapsto [0,1]  \\
f_{XYZ}  &= f_{post,XYZ} \circ f_{Q,XYZ} \circ f_{pre} : \mathbb{R}^6 \mapsto [0,1]  
\end{aligned}
\end{equation}
where $f_{Q,\cdot}$ is the quantum part.

The parametrized quantum circuit uses angle encoding with Pauli matrices, yielding an integer-valued, equidistant spectrum, and measures the expectation values of the Pauli-Z observable locally on each qubit, resulting in a $d$-dimensional output.
This results in the following Fourier spectra:
\begin{itemize}
\item $f^8_{ZZXY}: \mathbb{R}^8 \rightarrow \mathbb{R}$, with 2 encoding layers, i.e.,  $\Omega_{ZZXY} = [-2,-1,0,1,2]^8$ with spectrum size $\vert \Omega_{ZZXY} \vert = 390\,625$,
\item $f^8_{XYZ}: \mathbb{R}^8 \rightarrow \mathbb{R}$ with 5 encoding layers, i.e.,  $\Omega_{XYZ} = [-5,-4,\dots, 4,5]^8$, $\vert \Omega_{XYZ} \vert = 214\,358\,881$,
\item $f^6_{ZZXY}: \mathbb{R}^6 \rightarrow \mathbb{R}$  with 2 encoding layers, i.e., $\Omega_{ZZXY} = [-2,-1,0,1,2]^6$, $\vert \Omega_{ZZXY} \vert = 15\,625$,
\item $f^6_{XYZ}: \mathbb{R}^6 \rightarrow \mathbb{R}$, with 4 encoding layers, i.e., $\Omega_{XYZ} = [-4,-3,\dots, 3,4]^6$, $\vert \Omega_{XYZ} \vert = 531\,441$.
\end{itemize}
We use the optimal parameters provided by \cite{Pastori.14.02.2025}, one set of parameters trained in an idealized setting and another trained with shot noise and variance regularization. The training and test data used in \cite{Pastori.14.02.2025} are based on the  DYAMOND project \cite{Duras.2021,Stephan.2022,Stevens.2019} coarse-grained to a lower resolution following \cite{Grundner.2022,Grundner.2024}.

\section{Methods} \label{sec:methods}

As mentioned before, our aim is shadowing methods that require no training or regression stage and is instead as constructive as possible.  In our work, we focus on methods to bypass any training or regression stage in our shadowing process, which would allow replacing the input-output of the quantum circuit by the training data.

\begin{figure}[t]
\begin{center}
\includegraphics[width = 0.75\linewidth, trim={5cm 1cm 5cm 1cm},clip]{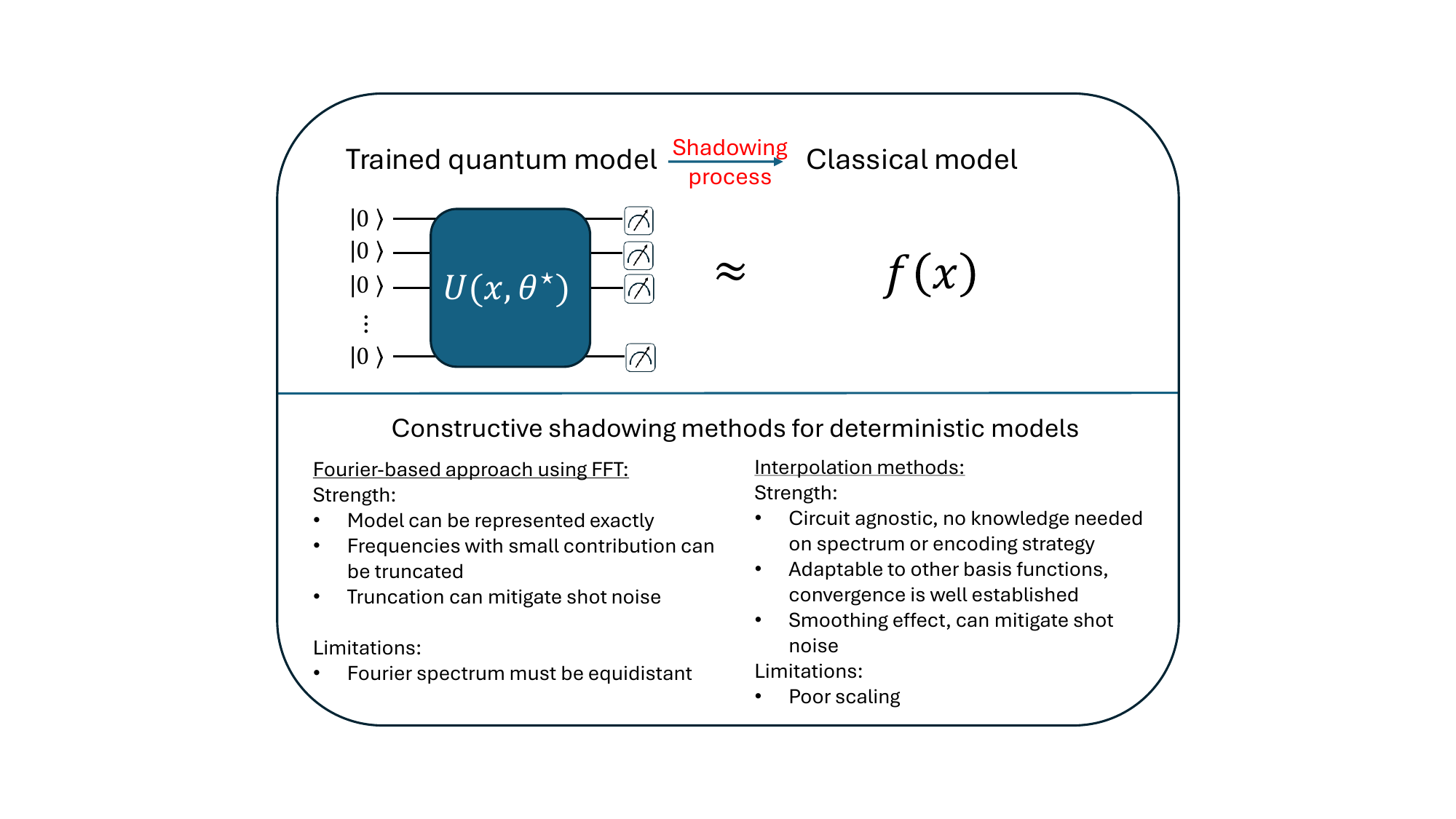}
\caption{Overview of constructive shadowing methods used in the work. The Fourier-based approach allows for an efficient, exact representation of the circuit, assuming that the Fourier spectrum is equidistant. As a circuit agnostic approach, which needs no knowledge of the encoding strategy, a quasi-interpolation is considered. Both methods can mitigate finite sampling noise under suitable conditions.} \label{fig:overview}
\end{center}
\end{figure}

For now, we focus on the application at hand, which is a shadowing process for the smaller models as in \cite{Pastori.14.02.2025}, with the aim to recover the function in the sense of achieving a comparable mean squared error and for shadowing methods where the quantum circuit is used explicitly and cannot be easily replaced by the training data.
An overview of the methods considered in this work is shown in Figure \ref{fig:overview}.

Since the encoding strategy in \cite{Pastori.14.02.2025} allows for an easy recovery of the coefficients of the Fourier series, our first approach for a shadowing procedure is the quantum Fourier model in Section \ref{sec:methods:qfm}. The Pauli-encoding in \cite{Pastori.14.02.2025} allows to use the fast Fourier transform to efficiently compute the coefficients of the entire spectrum. While there were some experiments done on approximating the Fourier model with distinct sampling of frequencies as in \cite{Hernicht.08.08.2025, JonasLandman.}, we do not report on them here, since they were not competitive with the methods investigated in this work.

One limiting factor of this approach is, of course, the size. While the FFT is an efficient algorithm to solve this problem, already for the largest of the models considered here  with eight input features with over 200 million frequencies it becomes impractical.
 If more sophisticated encoding strategies are used as suggested in \cite{Mitarai.2018,Liao.2022,PhysRevA.109.042421} and the spectrum is no longer equidistant, one needs to use methods for the non-uniform discrete Fourier transform, which do not always yield an efficient inverse \cite{Wiedmann.05.11.2024}. 

In Section \ref{sec:methods:interpolation} we consider an alternative, circuit-agnostic approach.
Since the functions we want to approximate are deterministic (discounting the effect of finite sampling noise), our problem is simply a quasi-interpolation problem, i.e., finding a suitable (finite) set of basis functions  $\{\phi_i \}_{i \in \{1, \dots N \}}$, $\phi_i : \mathbb{R}^d \mapsto \mathbb{R}$, and coefficients $\alpha_i \in \mathbb{R}$, such that
\begin{equation}
f(\bm{x}) \approx \sum_{i = 1}^{N} \alpha_i \phi_i(\bm{x}).
\end{equation}

In general, an interpolation problem can be solved by a system of linear equations or a least-squares problem as in Eq.~\eqref{eq:least_square}, which is inefficient  to solve if $N$ is large even with stochastic methods. 
We use that we can evaluate the function $f$ at arbitrary points. By choosing appropriate basis functions $\phi_i$ and support points $\bm{z}_i$, we can ensure that our system of linear equations becomes easier to solve. In the easiest case,  we can approximate
\begin{equation} \label{eq:interpolation}
f(\bm{x}) \approx \sum_{i=1}^N f(\bm{z}_i) \phi_i(\bm{x}).
\end{equation}
This is well researched in numerical analysis, and reliable convergence results are available for a variety of basis functions like splines and piecewise polynomial functions, see e.g. \cite{Suli.2012} for an introduction. However, classical numerical methods are prone to suffer from the curse of dimensionality. Already for the six dimensional problem, the interpolation methods have limited practicability.

Nevertheless, we opt to explore this approach, since it is the most straightforward, constructive method to reconstruct a given function, especially if a circuit architecture-agnostic method is needed. This could, for example, be the case if the eigenvalues of the encoding matrices are unknown or inefficient to compute. Additional motivation for this strategy is, that while we know that the Fourier basis describes our circuit, we actually have no guarantee that it is suitable to approximate the underlying data set well, especially since the QML models have no trainable parameters in the encoding layers to adjust the frequencies to better represent the data.

 Another issue presents itself if one wants to create a shadowing process for stochastic models. The interpolation methods in this work rely on the deterministic nature of the function which we want to shadow. A previous publication \cite{GarciaMartin.2025} shows that Gaussian Processes can be used, which might be interesting when considering stochastic models.
 
Since we evaluate the shadowing methods here only for a single application -- the cloud cover parameterization -- evaluating the shadowing procedures with respect to Definition \ref{def:shadow}  (cf. \cite{Jerbi.2024}) is difficult. 
The shadowing procedures used in this work all require the evaluation of the quantum model on a grid. For the quantum Fourier model the quantum model needs to be evaluated on a grid on $[0,2\pi]^d$ of size $(2L+1)^d$. For the interpolation methods the evaluation of a grid the size of order $(2^{J}+1)^d$ for a refinement parameter $J$ is necessary.  While the effect of the refinement parameter $J$ on the interpolation error depends on the interpolation strategy used, the dependence of the grid size on $d$ suggests that these methods scale too poorly to be considered viable shadowing processes. 
Another aspect is that the shadowing process in Definition \ref{def:shadow} is evaluated by comparing it to the quantum model itself, not the data the quantum model was trained on. Therefore it is not taking into account any potential improvements the shadowing process might have on the overall problem compared to the quantum model.

\subsection{Approximation using the Fourier model} \label{sec:methods:qfm}

First, we consider the approximation using the quantum Fourier model. Here, the shadowing process involves only approximating the quantum circuit, i.e., the function $f_{Q,\cdot}: \mathbb{R}^d \rightarrow \mathbb{R}^d$ of Eq.~\eqref{eq:qml_clc_arch} where the argument $(\cdot)$ acts as a place holder for the circuit architectures $XYZ$ and $ZZXY$.
 The complete shadow model is then constructed by replacing $f_{Q}$ by $\widetilde{f}_Q$, i.e.,  
\begin{equation}
\widetilde{f}_{\cdot} = f_{post,\cdot} \circ \widetilde{f}_{Q,\cdot} \circ f_{pre} \, .
\end{equation}

The frequency spectrum has the form $\Omega = [-L,\cdots, L]^d$, where $L$ is the number of encoding layers and $d$ the number of qubits, corresponding to the number of input features. Using the QFM
\begin{equation}
\widehat{f}_Q(\bm{x}) = \sum_{\bm{\omega} \in \Omega} c_\omega e^{-i \bm{x} \cdot \bm{\omega}}
\end{equation}
with $\bm{c_\omega}\in \mathbb{R}^d$ to represent the quantum circuit, the shadowing procedure is just a question of computing the coefficients $\bm{c_\omega}$. In our case, due to the form of the spectrum, this can be done  using the inverse $d$-dimensional discrete Fourier transform
\begin{equation}
\bm{c_\omega } = \frac{1}{(2L+1)^d}\sum_{\bm{k} \in \Omega} f_Q\left( \frac{2\pi}{2L+1}\bm{k}\right)e^{\frac{2\pi}{2L+1}(\bm{\omega}\cdot \bm{k})},
\end{equation}
where we evaluate the quantum circuit $f_Q$ at grid points corresponding to the spectrum scaled to $[0,2\pi]^d$. 
Since $f_Q$ is real valued, we can use the symmetry 
\begin{equation}\label{eq:sym_coeff}
\bm{c_{\omega}} = \bm{c_{-\omega}}^\star
\end{equation} and alternatively write 
\begin{equation}
f_Q(\bm{x}) = \sum_{\bm{\omega }\in \Omega^+} a_\omega \cos(\bm{x} \cdot \bm{\omega}) + b_\omega \sin(\bm{x} \cdot \bm{\omega})
\end{equation}
with $\bm{a_\omega} = \bm{c_\omega} + \bm{c_{-\omega}}$, $\bm{b_\omega} = -1/i(\bm{c_\omega}-\bm{c_{-\omega}})$ and $\Omega^+ \subset \Omega$, as described in Section \ref{sec:background:qfm}. %

This allows to reconstruct the entire circuit by evaluating the quantum circuit at $(2L+1)^d$ points and using the FFT, which is well known to scale $O(N \log (N))$ where $N = (2L+1)^d $. This approach was also used in \cite{Fontana.11.08.2022,Fontana.17.06.2022,Strobl.2025,Wiedmann.05.11.2024}. \\
We reduce the size of the shadow model by choosing a subset $\Omega_{M} \subset \Omega$ of the $M$ frequencies with the largest coefficient (largest with respect to the Euclidean norm $\Vert \cdot \Vert$). Due to the symmetry \eqref{eq:sym_coeff} we have $\Vert \bm{c_{\omega} }\Vert = \Vert \bm{c_{-\omega}} \Vert$, such that we can ensure that the truncated series is still real by choosing an even $M$. Some care has to be taken about the zero frequency. For an easier implementation we ensure that the zero frequency vector is always included in $\Omega_M$. 
We remark that in $\Omega_{+,M}$, only $M/2+1$ frequencies are left after truncation, however with two sets of coefficients $\bm{a_\omega, b_\omega}$ (with the exception for the zero frequency, for which the sine term in Eq.~\eqref{eq:sym_coeff} vanishes), such that the degree of freedom remains the same.

 In an idealized setting with infinite shots and no sampling noise, this leads to a truncation error, while with increasing $M$ we can recover the entire function. The influence of the number of kept frequencies $M$ on the truncation error depends on how much the function concentrates within the spectrum. 
This is similar to the observation from \cite{RyanSweke.}, where classical simulatibility of the inference stage depends on how the function concentrates and the ability to determine where it concentrates.
Unfortunately, here we can only determine this after computing all coefficients for the entire spectrum, and evaluating the shadow model on a sufficiently large test set.

\subsection{Approximation using quasi interpolation methods}\label{sec:methods:interpolation}

We consider a quasi-interpolation approach with basis functions constructed with tensor products of piecewise affine linear functions following \parencite{Bungartz.2004} on an equidistant grid $\widehat{X} = \{j\pi 2^{-J+1} \}_{j \in \{0, \dots 2^{J} \}}^d $ on $[0,2\pi]^d$ for a refinement parameter $J \in \mathbb{N}$. The basis functions are given by
\begin{equation*}
\Phi_j(\bm{x}) = \prod_{k = 1}^d \phi\left(\frac{\bm{x}_k-\widehat{\bm{x}}_{j,k}}{h}\right) \text{ for }  \widehat{\bm{x}}_j \in \widehat{X},j \in \{0, \dots 2^J \}^d \text{ with } \phi(x) = \chi_{x \in [-1,1]} \max(0,1- \vert x\vert).
\end{equation*}
This method has the advantage of being straightforward to implement with
\begin{equation}\label{eq:qi_full}
f(\bm{x}) \approx \sum_{j \in \{0, \dots 2^{J} \}^d} f(\widehat{\bm{x}}_j) \Phi_j(\bm{x}).
\end{equation}
To generate a shadow model we interpolate the function  $f_{post, \cdot} \circ f_{Q,\cdot}: [0,2\pi]^d \mapsto \mathbb{R}$, instead of just the quantum part $f_{Q,\cdot}$ as in the previous section. 
The obvious draw back to this shadowing process is that the number of grid points grows exponentially with the refinement parameter $J$ and the number of dimensions $d.$

A way to circumvent the exponential growth of the number of grid points for high-dimensional interpolation is using sparse grids as in \cite{Kempf.2023,Griebel.2025,Kolomoitsev.2023,Hubbert.2023,Bungartz.2004,Gao.2024}, which promise for $O(N \log(N)^{d-1})$ grid points an interpolation error $O(N^{-2} \log(N)^{d-1})$ with respect to the $L^2$-norm for a refinement parameter $N$ \parencite{Bungartz.2004}. However, an investigation for the cloud cover model has shown no improvements. Details can be found in  Appendix \ref{app:sparse_grids}.

We forgo a detailed error analysis here for the interpolation methods on both the full and the sparse grids and instead refer to the error estimates in \parencite{Bungartz.2004} for the given basis functions.

\subsection{Incorporating finite sampling noise}\label{sec:methods_shots}

So far we have considered an idealized setting, where we can evaluate the function $f_Q$ without any noise. Since an evaluation of results obtained on a  quantum computer involves the noise of finite measurements, this needs to be reflected in the definition of a shadow model. We denote $N_s$ as the number of shots (or samples) that we take for each evaluation of the circuit as a specific data-point, and denote the approximation of the function $f$ as $f^{N_s}$. For example, when constructing the shadow model using the QFM, we now compute the noisy coefficients
\begin{equation}
c^{N_s}_\omega = \frac{1}{(2L+1)^d} \sum_{\bm{k}\in \Omega} f^{N_s}_Q\left(\frac{2\pi}{2L+1}\bm{k} \right)e^{i\frac{2\pi}{2L+1}\bm{\omega \cdot k}}
\end{equation}
and the corresponding partial Fourier series
\begin{equation}
\widetilde{f}_Q^{N_s}(\bm{x}) = \sum_{\bm{\omega }\in \Omega}c^{N_s}_\omega e^{-i\bm{x \cdot \omega}}.
\end{equation}
The function $\widetilde{f}^{N_s}$ itself is deterministic, but depends on the samples taken under the influence of shot noise, which can be seen as drawing a sample from a space of admissible functions. 
There are several publications on the effect noise has on the QFM, e.g.,  \cite{Fontana.2025,Lai.2026,Fontana.17.06.2022,Franz.11.06.2025}, however, mostly concerning the effect of hardware noise other than finite sampling noise.
A simplified discussion on the influence of noise on the truncated quantum Fourier model and the interpolation can be found in Appendix \ref{app:noise}.

 \section{Results} \label{sec:results}

Next, we discuss the performance of the shadow models compared to the QML models for cloud cover parametrization. 
In the following sections, we use three different test data sets $\mathcal{D}_c \subset \mathcal{D}_r \subset \mathcal{D} \subset \mathcal{D}_0$.
The test data set $\mathcal{D}_0$ is a subset of the test data set used in \cite{Pastori.14.02.2025} based on the DYAMOND project \cite{Duras.2021,Stephan.2022,Stevens.2019} coarse-grained to an approx. 80 km resolution following \cite{Grundner.2022,Grundner.2024}. From it, we removed all cells for which the parametrization does not need to be called because the cloud water and cloud ice content is too small for clouds to occur. The remaining data set $\mathcal{D}$ makes up around 37 percent of $\mathcal{D}_0$ and consists of approximately $9\cdot 10^6$  cells. For validations that take more resources, like the interpolation methods or hardware experiments, we use a reduced test data set $\mathcal{D}_r$ of size $4000$ cells. The interpolation on the sparse grid was only evaluated on a test set for cirrus clouds $\mathcal{D}_c$ of the size $1000$ cells. While these are rather small data sets, we expect that the approximation results would hold also for larger test sets, considering that this is true when comparing the performance of the interpolation methods on $\mathcal{D}_c$ and $\mathcal{D}_r$ in Section \ref{sec:results:interpolation}, and the performance of the Fourier-based shadow model in particular in Section \ref{sec:results:qfm:iqm}.

In Section \ref{sec:results:interpolation} we discuss the shadowing process using the interpolation techniques described in Section \ref{sec:methods:interpolation}. In Section \ref{sec:results:qfm} we consider the shadow models constructed with the truncated quantum Fourier model.
The experiments in Sections \ref{sec:results:interpolation} and \ref{sec:results:qfm:sv} are state vector simulations using PennyLane's \texttt{default.qubit} \cite{Bergholm.2018}. Section \ref{sec:results:qfm:iqm} tests the shadowing process and the QML models on the superconducting quantum computer Euro-Q-Exa. 

\subsection{Approximation using piece-wise affine linear quasi interpolation}\label{sec:results:interpolation}
In this section, we show the results of the shadow models using the interpolation method as described in Section \ref{sec:methods:interpolation}.
We compare here the mean squared errors of the test data set for the quasi-interpolation operator \eqref{eq:qi_full}, denoted by $\mathcal{I}f_\cdot$ on the full grid to the mean squared error of the respective 6-qubit QML models (denoted by $f^6_{XZY}$ and $f^6_{ZZXY}$).

\begin{figure}
\includegraphics[width = 0.5\linewidth]{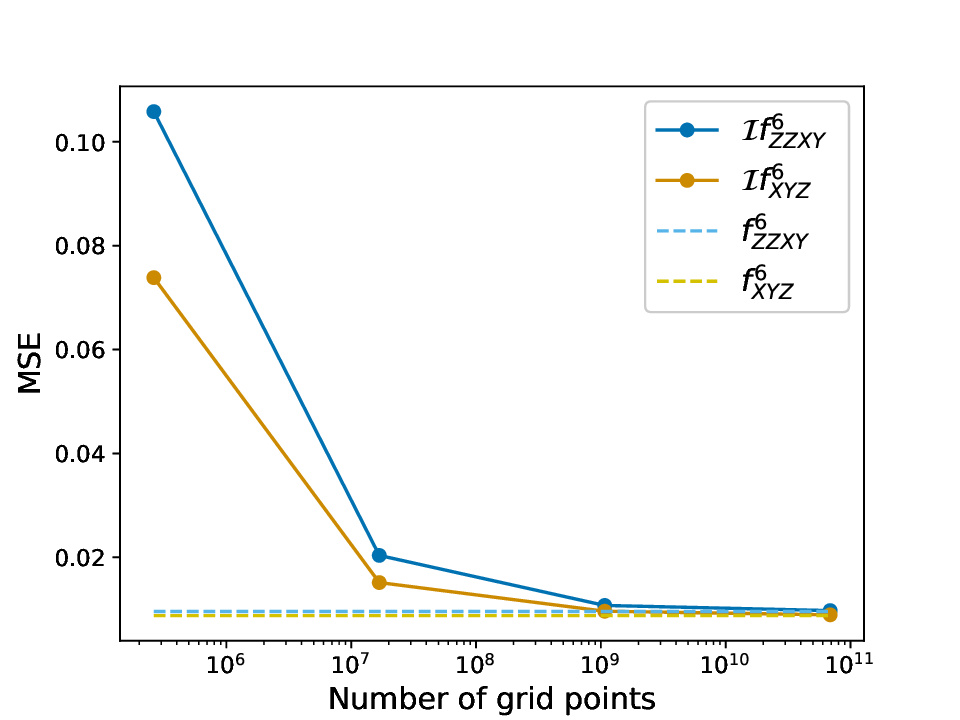}
\includegraphics[width = 0.5\linewidth]{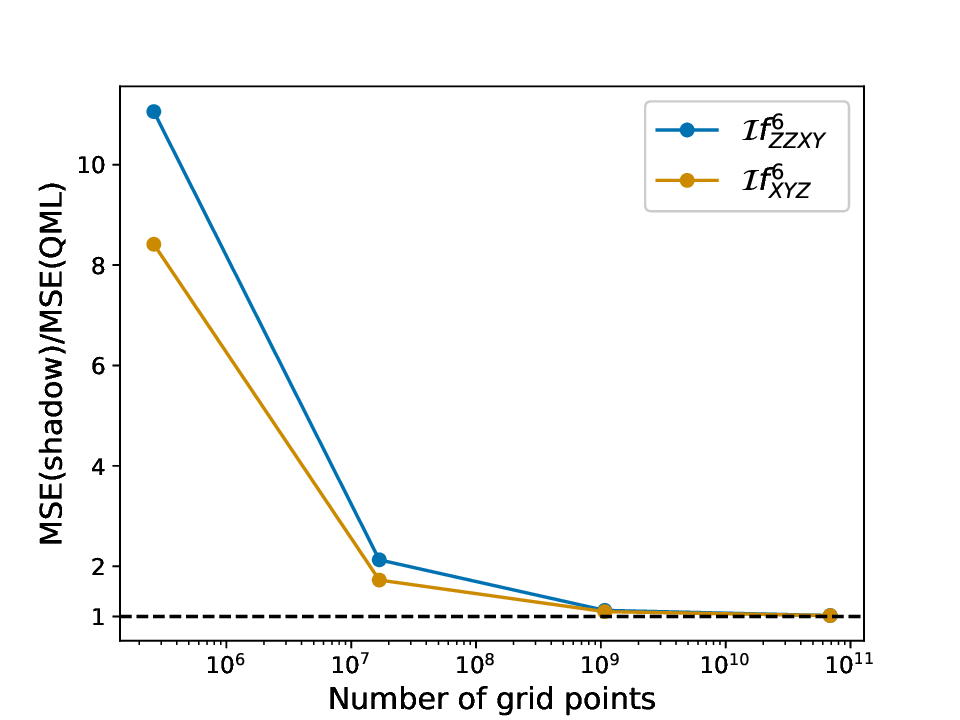}
\caption{Left: MSE of the QML models $f^6_{ZZXY},f^6_{XYZ}$ on the reduced test data set $\mathcal{D}_r$ (dashed lines) compared to MSE of interpolants on full grids $\mathcal{I}f^6_{ZZXY},\mathcal{I}f^6_{XYZ}$ (solid lines) plotted against the number of points in the grid we interpolate on. The interpolants are defined on grids with refinement parameters $L \in \{3,4,5,6\}$, indicated by the dots. Right: Ratio of the MSE of interpolants and MSE of QML models, $\frac{MSE(\mathcal{I}f_{\cdot})}{MSE(f_{\cdot})}$ as a function of the number of grid points. The closer the fraction is to the value 1 (indicated by the dashed line as guide to the eye), the better the shadow model approximates the QML model.  QML models were trained and evaluated in a noiseless regime.}\label{fig:qi_noiseless}%
\end{figure}

 First, in Figure \ref{fig:qi_noiseless}, we consider a noiseless setting, where both the training and evaluation of the QML model is done without shot noise. We compute the interpolants on the grids $ \{j\pi 2^{-J+1} \}_{j \in \{0, \dots 2^{J} \}}^d $, for $L \in \{3,4,5,6\}$. The figure shows the MSE of the interpolants plotted against the total number of points in the grids. For the quasi interpolation on the full grid, we observe a convergence rate of approximately $\mathcal{O}(N^{-0.5})$ for $N$ grid points. However the number of required grid points significantly limits the practicability of this approach. 
 We also investigated the use of sparse grids in Appendix \ref{app:sparse_grids}. However, since this showed no improvements, we did not investigate it further.

\begin{figure}
\includegraphics[width = 0.5\linewidth]{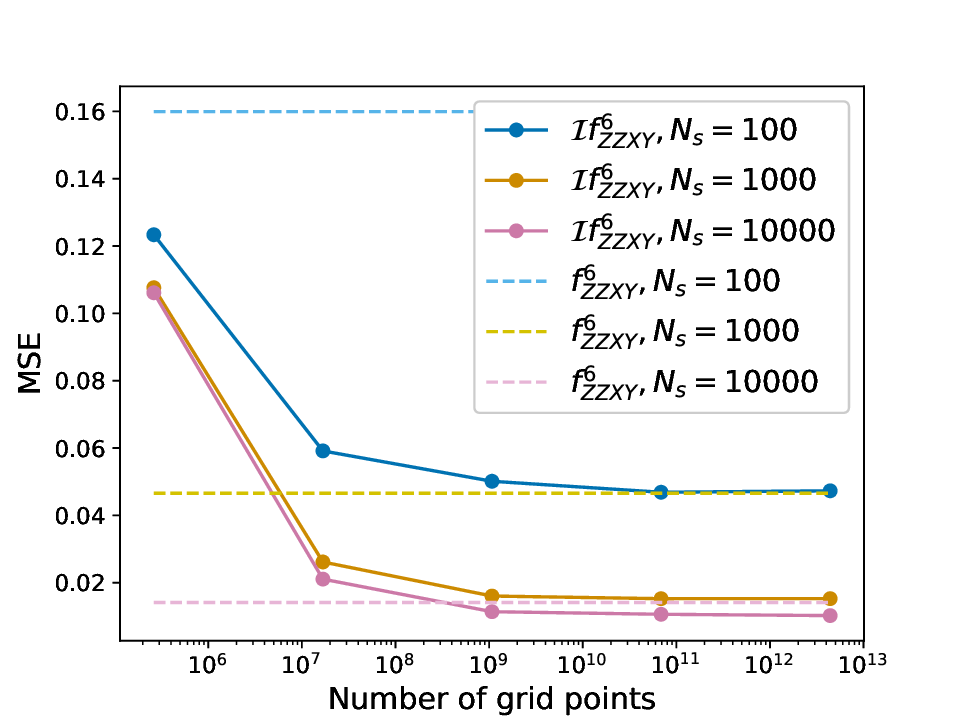} 
\includegraphics[width = 0.5\linewidth]{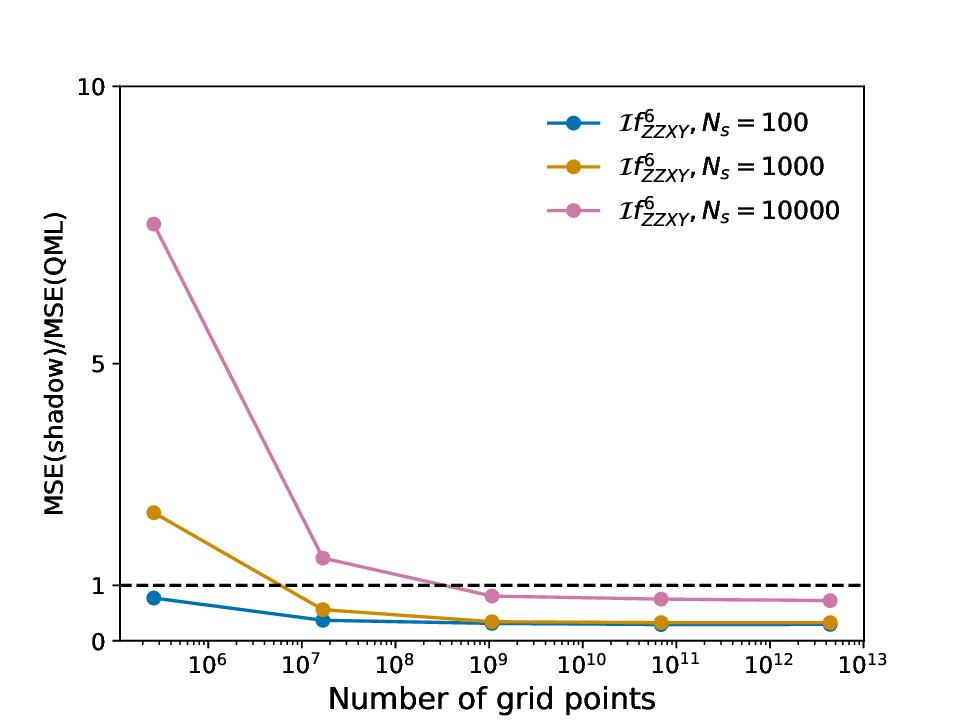}
\includegraphics[width = 0.5\linewidth]{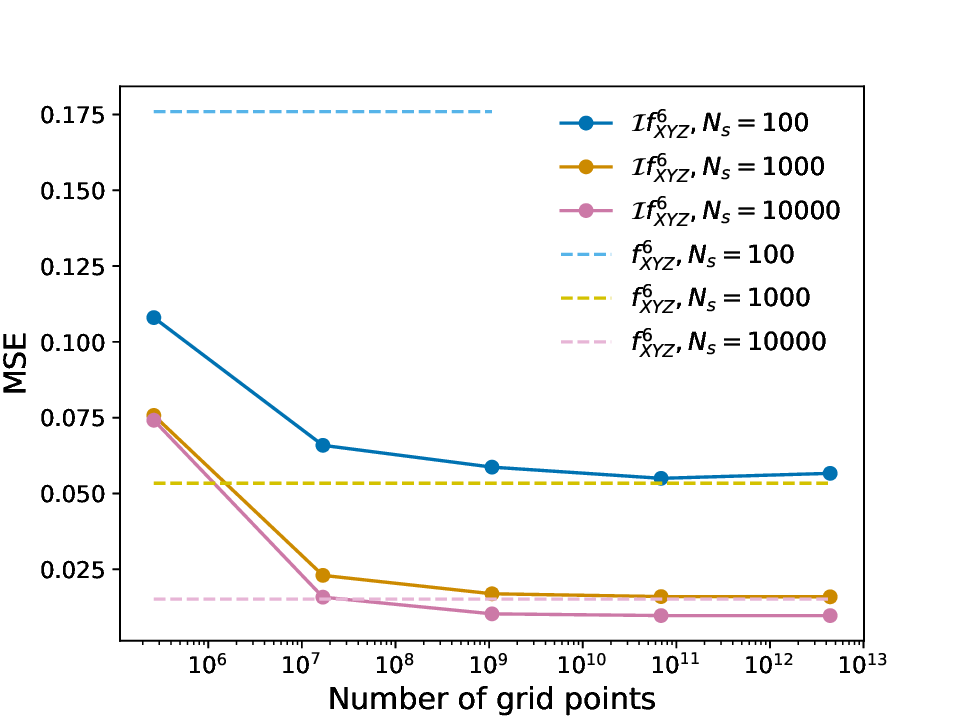} 
\includegraphics[width = 0.5\linewidth]{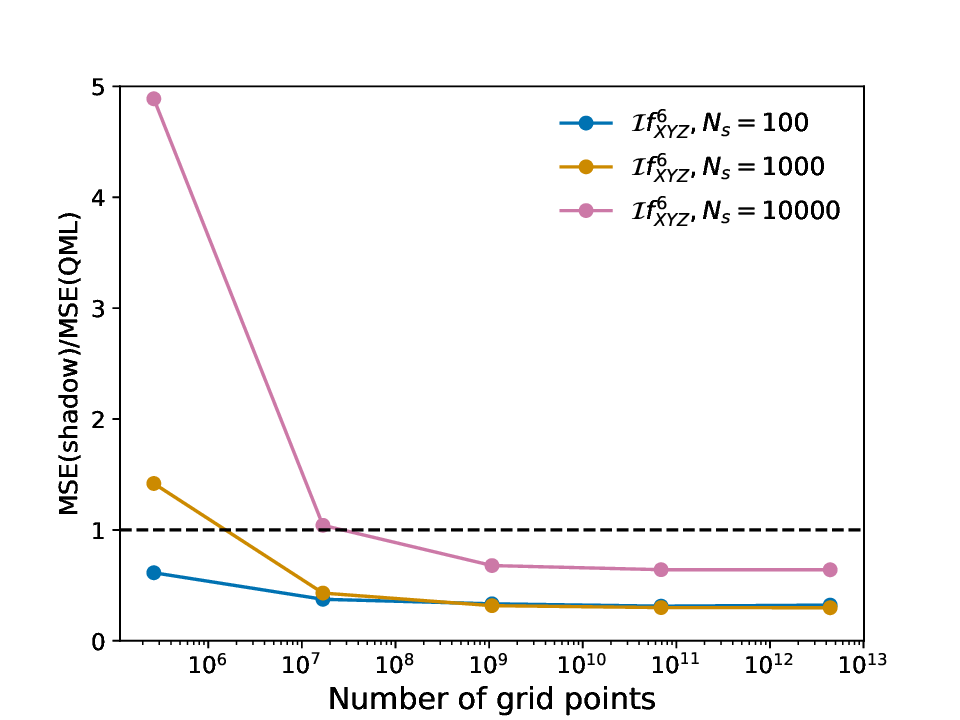}
\caption{Comparison of interpolants on the full grid to QML models evaluated with shot noise on the reduced test data set $\mathcal{D}_r$ for the $ZZXY$ architecture (top row) resp. $XYZ$ architecture (bottom row). Left: MSE of interpolants (solid lines) compared to MSE of QML model (dashed lines) evaluated with 100 (blue), 1000 (yellow) and 10\,000 (pink) shots.  The interpolants are defined on grids with refinement parameters $L \in \{3,4,5,6,7\}$, indicated by the dots. Right: Ratio of the MSE of noisy evaluations of the interpolants against the MSE of the noisy evaluation of the QML model, i.e. $\frac{MSE(\mathcal{I}f^{N_s}_{\cdot})}{MSE(f_{\cdot}^{N_s})}$ (solid lines), as a function of the number of grid points. The closer the fraction is to the value 1 (indicated by the black dashed line as guide to the eye), the better the shadow model approximates the noiseless QML model. A value below 1 indicates that the interpolants outperforms the QML models with the respective number of shots used. The interpolation mitigates the effect of finite sampling noise. The models were trained in a noiseless regime.} \label{fig:qi_noisy}
\end{figure}

Next, we compare the MSE incorporating finite sampling noise. For Figure \ref{fig:qi_noisy}, we still consider the QML models trained in a noiseless regime, but evaluate them with shot noise for the computation of the coefficients. We plot the MSE of the interpolants and the QML model evaluated with shot noise for each architecture and additionally the difference in MSE to that of the models in the noiseless regime, cf. Figure \ref{fig:qi_noiseless}. Here, we consider the interpolants on grids with a refinement parameter $L\in \{3,4,5,6,7\}$. We observe that the interpolants yield a lower MSE compared to the QML model evaluated with shot noise in most cases, such that the interpolant-based shadow model has a lower MSE compared to the original QML model evaluated with the same number of shots as used to compute the coefficients for the shadow model. The more shots are used, the finer the grid needs to be to observe this effect. Another interesting observation, is that the interpolants with shot noise seem to converge to the QML model with 10 times the number of shots used yielding a comparable MSE. Our hypothesis is that this is tied to the choice of basis functions and their variance regulating effect as discussed in Appendix \ref{app:noise}, which would would be interesting to investigate more thoroughly in the future.

\begin{figure}
\includegraphics[width = 0.5\linewidth]{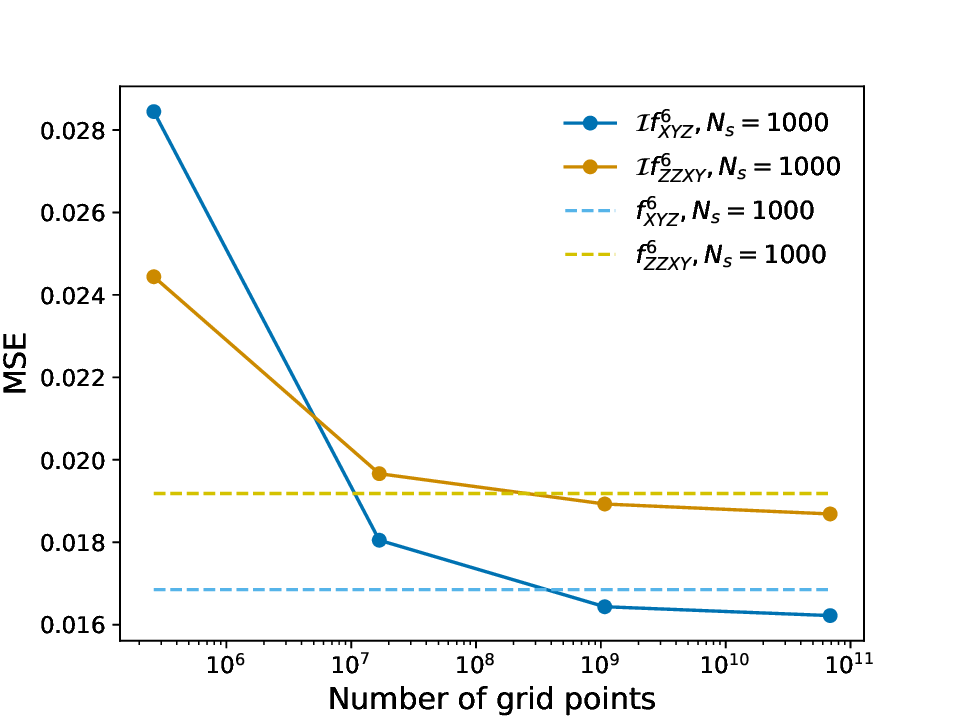}
\includegraphics[width = 0.5\linewidth]{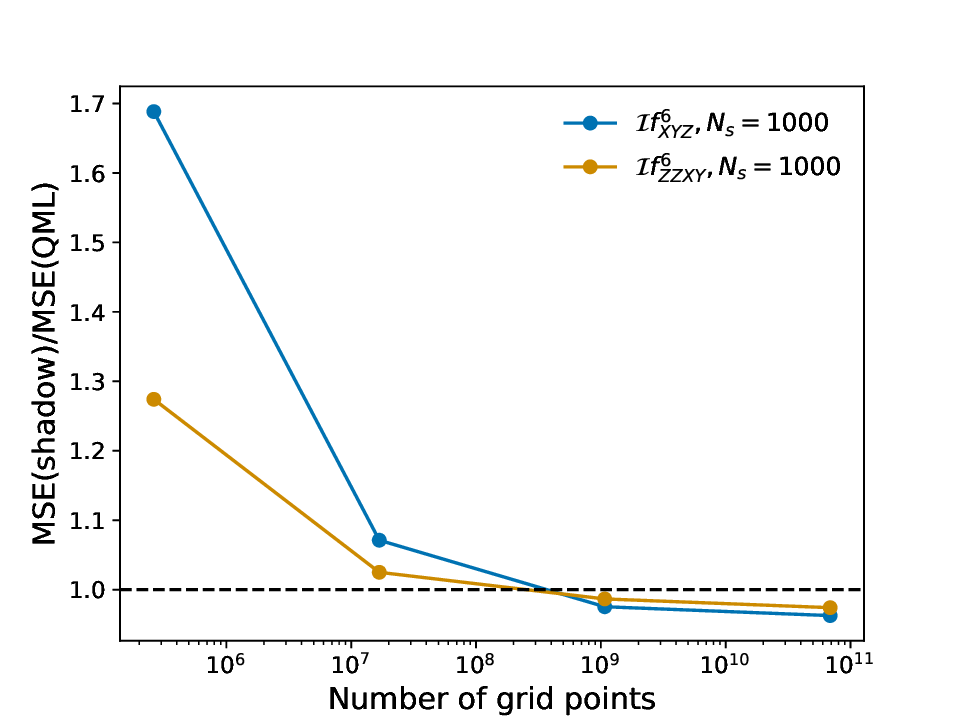}
\caption{Left: Comparison of interpolant of full grid (solid lines) to QML model (dashed lines) trained with shot noise (1000 shots) and variance regularization evaluated with shot noise on the reduced test data set $\mathcal{D}_r$, for the two architectures (XYZ, blue; ZZXY, orange). Right: Ratio of the MSE of the interpolants and the QML models, $\frac{MSE(\mathcal{I}f_{\cdot})}{MSE(f_{\cdot})}$ as a function of the number of grid point, both evaluated with  $N_s = 1000$ shots. The closer the fraction is to the value 1 (indicated by the black dashed line as guide to the eye), the better the shadow model approximates the noisy QML model.  A value below 1 means that the shadow model outperforms the QML model. The interpolation mitigates the effect of finite sampling noise for finer grids. The models were trained in a noisy regime with $N_s = 1000$.}\label{fig:qi_noisy_varreg}
\end{figure}

Lastly, we compare the same architectures, but with parameters trained in a noisy regime \cite[Section 5]{Pastori.14.02.2025}, trained with 1000 shots and variance regularization to mitigate the impact of sampling noise during training, interpolated on grids with $L \in \{3,4,5,6\}$.  In Figure \ref{fig:qi_noisy_varreg} we compare the shadow model with coefficients computed from the QML models with 1000 shots to the MSE of the QML models evaluated with 1000 shots. We still observe the error mitigation effects, but lessened and only on the finer grids.

\subsection{Approximation using the Fourier model} \label{sec:results:qfm}
This section discusses the performance of the shadowing process using the truncation of the quantum Fourier model as described in Section \ref{sec:methods:qfm}. 
We first consider the simulated experiments with and without finite sampling noise. In the last section, we consider shadow models generated using the quantum computer Euro-Q-Exa.

\subsubsection{Simulated experiments} \label{sec:results:qfm:sv}

\begin{figure}[h]
\begin{center}
\includegraphics[width = 7.5cm]{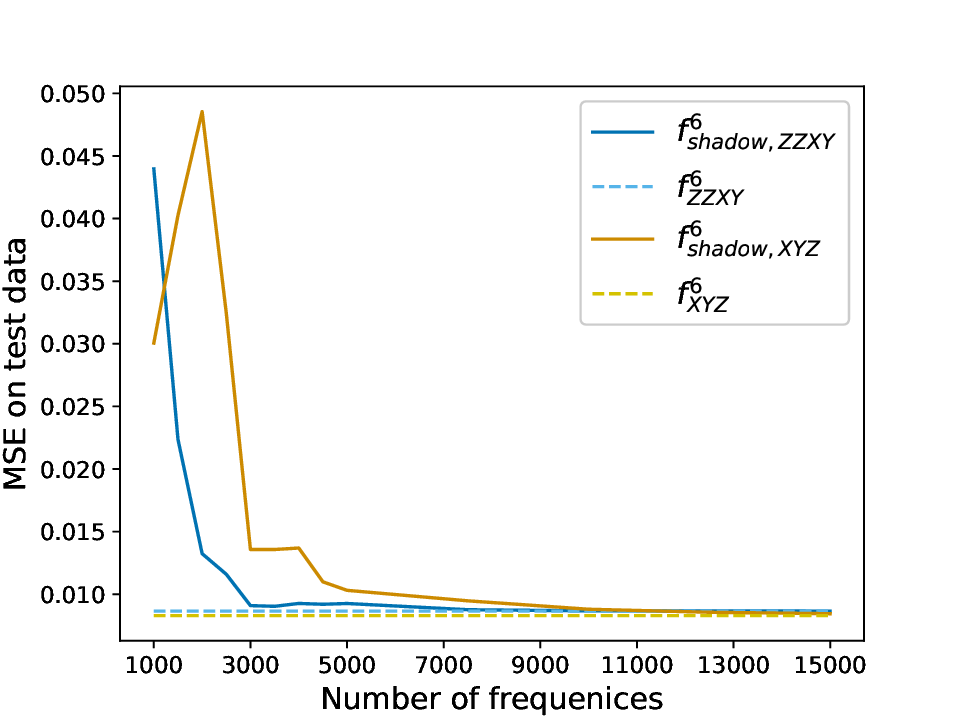}
\includegraphics[width = 7.5cm]{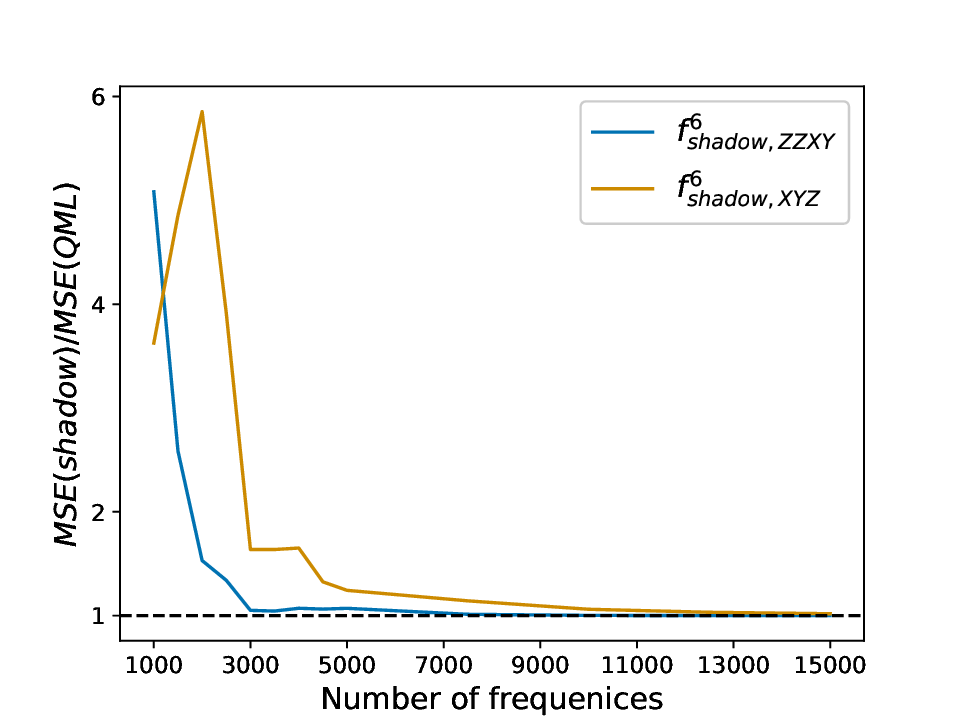}
\caption{Left: MSE of approximation of $f^6_{ZZXY}$ (blue), $f^6_{XYZ}$ (yellow) using the quantum Fourier model (solid lines) as function of number of retained frequencies (x-axis) compared to the MSE of the QML models evaluated with an infinite number of shots (dashed lines). Right: Ratio of the MSE of the truncated Fourier models and the QML models, $\frac{MSE(f_{shadow, \cdot})}{MSE(f_{\cdot})}$ as a function of kept frequencies. The closer the fraction is to the value 1 (indicated by the black dashed line as guide to the eye), the better the shadow model approximates the noisy QML models, evaluated on test data set $\mathcal{D}$. 
Around 7500 to 12500 frequencies seem to be enough to approximate the QML models $f_{ZZXY}$ and $f_{XYZ}$ respectively. The models were trained and evaluated in a noiseless regime.} \label{fig:fft_infshots}
\end{center} 
\end{figure}

First, we compute the shadow model using the quantum Fourier model, where we evaluate the QML model without any finite sampling noise, i.e., assuming an infinite number of shots. In this section, we evaluate both the shadow models and the QML models on the entire test data set $\mathcal{D}$. Figure \ref{fig:fft_infshots} shows the MSE of the truncated Fourier-based shadow models compared to the MSE of the QML models on the entire test data set as a function of the retained frequencies $M$, as described in Section \ref{sec:methods:qfm}. Notably, only a fraction of the spectrum is necessary to approximate the circuit. For the smaller of the models, $f_{ZZXY}$, around 7500 of 15\,625 frequencies, i.e., less than a half of the spectrum, and for $f_{XYZ}$ around 12500 of 531\,441, i.e., less than 2,5 percent, suffices to approximate the QML model well enough in the sense that the ratio of the MSE of the shadow model to the QML model  satisfies
\begin{equation}\label{eq:frac:metric}
     \frac{MSE(f_{shadow, \cdot})}{MSE(f_{\cdot})}  < 1.05.
\end{equation}
To compare this to the performance of the interpolation methods, only the last refinement shown in Figure \ref{fig:qi_noiseless}, i.e., with around $6 \cdot 10^{10}$ grid points, approximates the QML models according to this metric.

\begin{figure}[h]
\includegraphics[width = 0.5\linewidth]{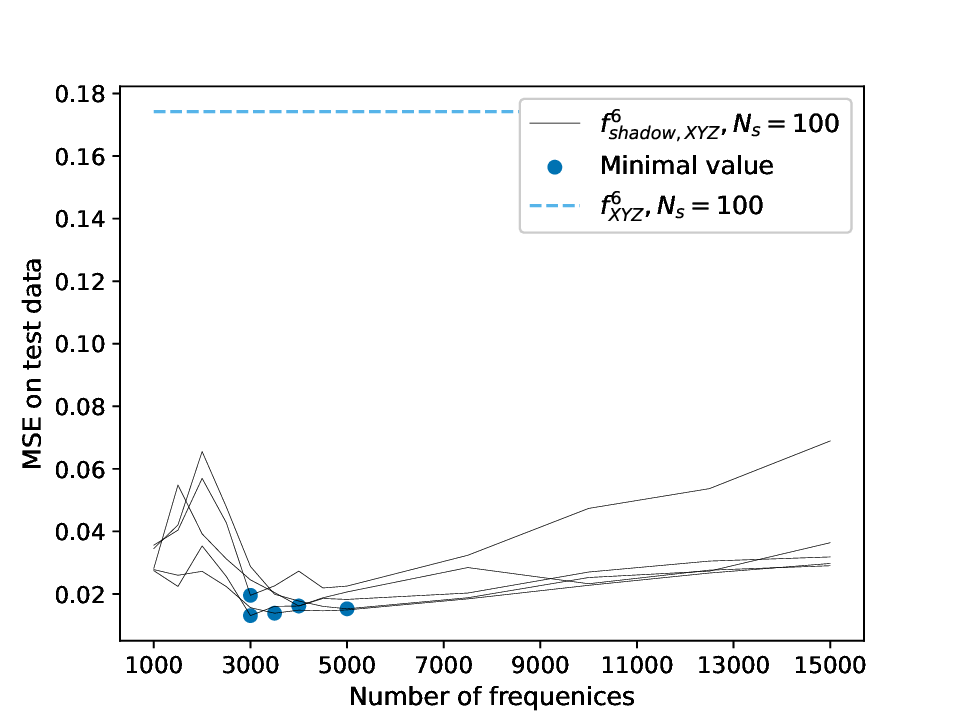}
\includegraphics[width = 0.5\linewidth]{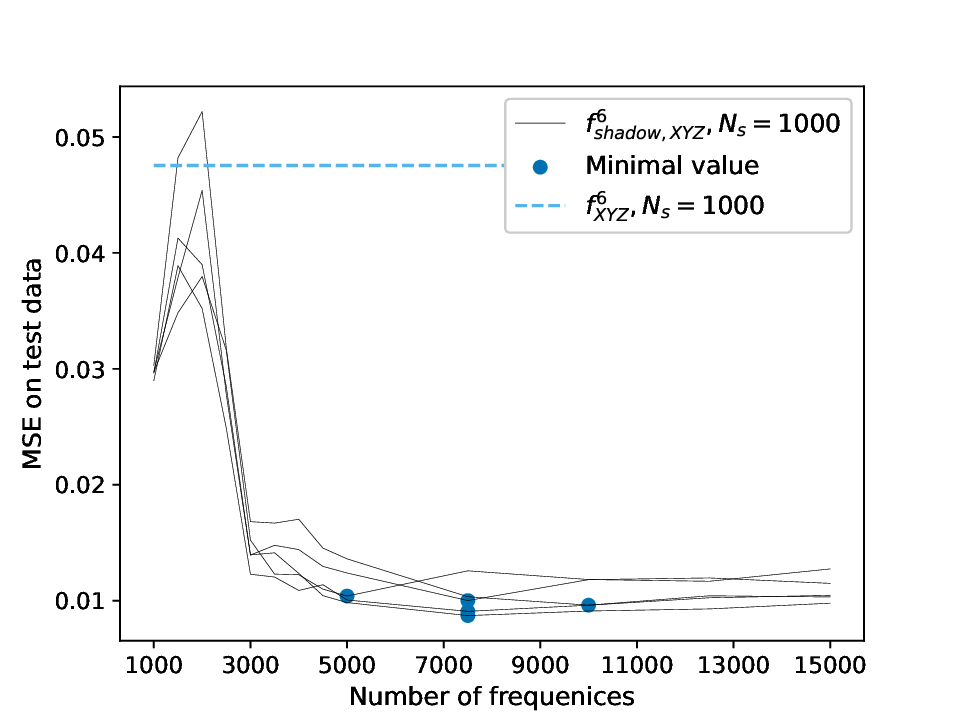} \\
\includegraphics[width = 0.5\linewidth]{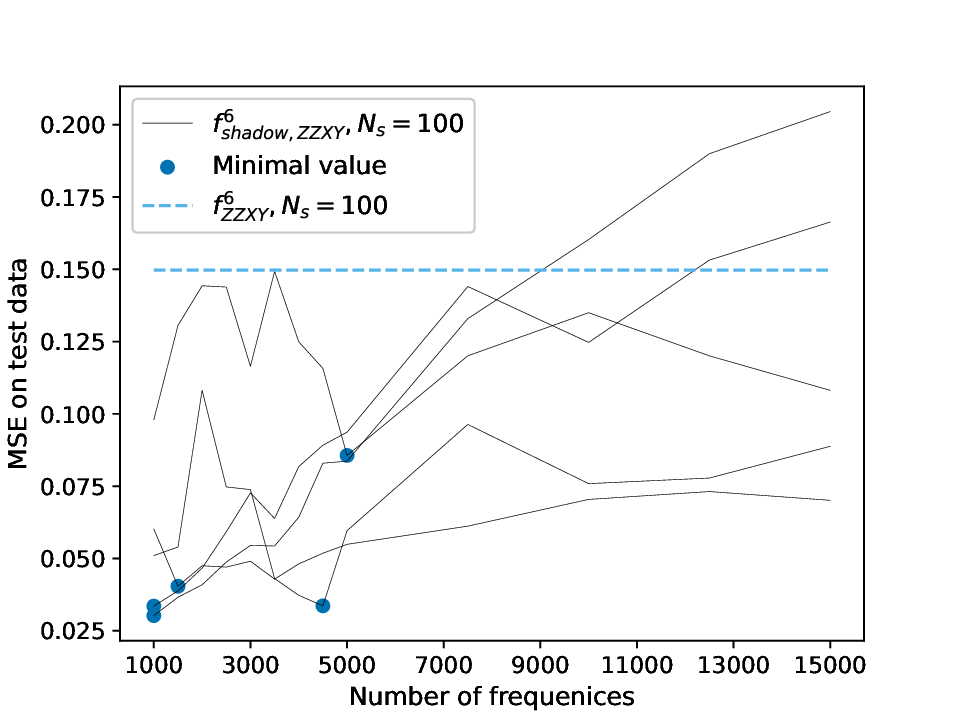}
\includegraphics[width = 0.5\linewidth]{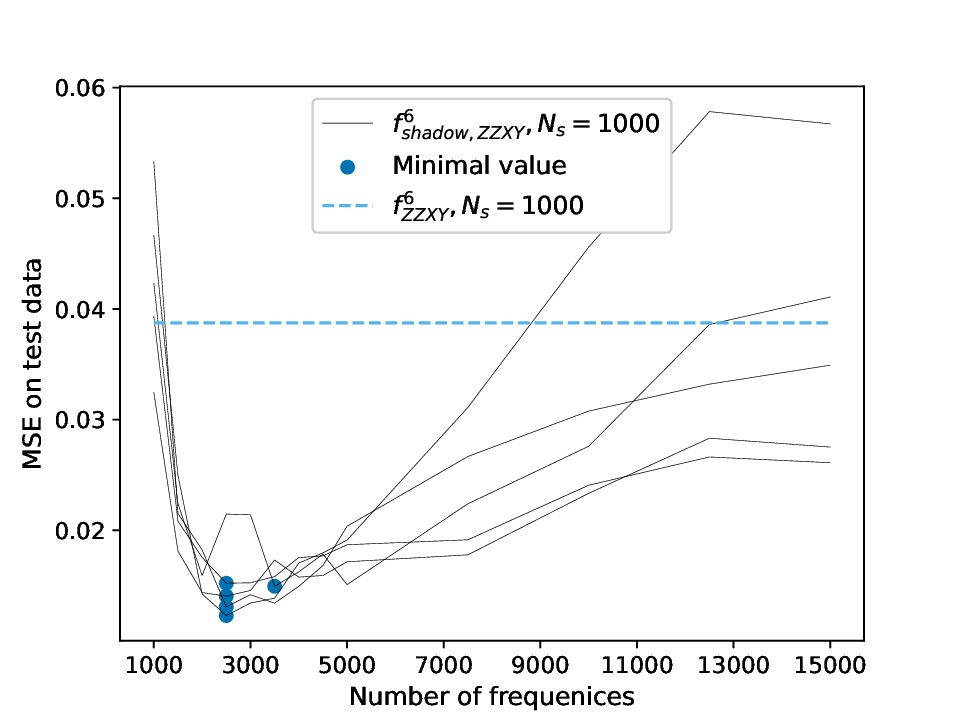} 
\caption{MSE on test data $\mathcal{D}$ of QML models compared to MSE of Fourier-based shadow models for the two architectures (top: XYZ, bottom: ZZXY) as functions of frequencies kept (x-axis). Infinite shots were used during training. To compute the coefficients of the QFM, 100 resp. 1000 (left/right column) shots were used in the evaluation of the circuit. Grey: Each line corresponds to a Fourier series as a function of retained frequencies (x-axis), with a distinct set of coefficients each computed using $N_s$ number of shots to account for the variability due to shot noise. Blue dashed line: MSE of QML model evaluated with 100 resp. 1000 shots on the test data set. Blue dots represent the minimum of the MSE of various shadow models. The truncation of the frequencies spectrum yields a lower MSE than a direct evaluation of the QML model.  Evaluated on test data set $\mathcal{D}$.}\label{fig:fft_inft_1000s}
\end{figure}

\begin{figure}[h]
\begin{center}
\includegraphics[width =  0.45\linewidth]{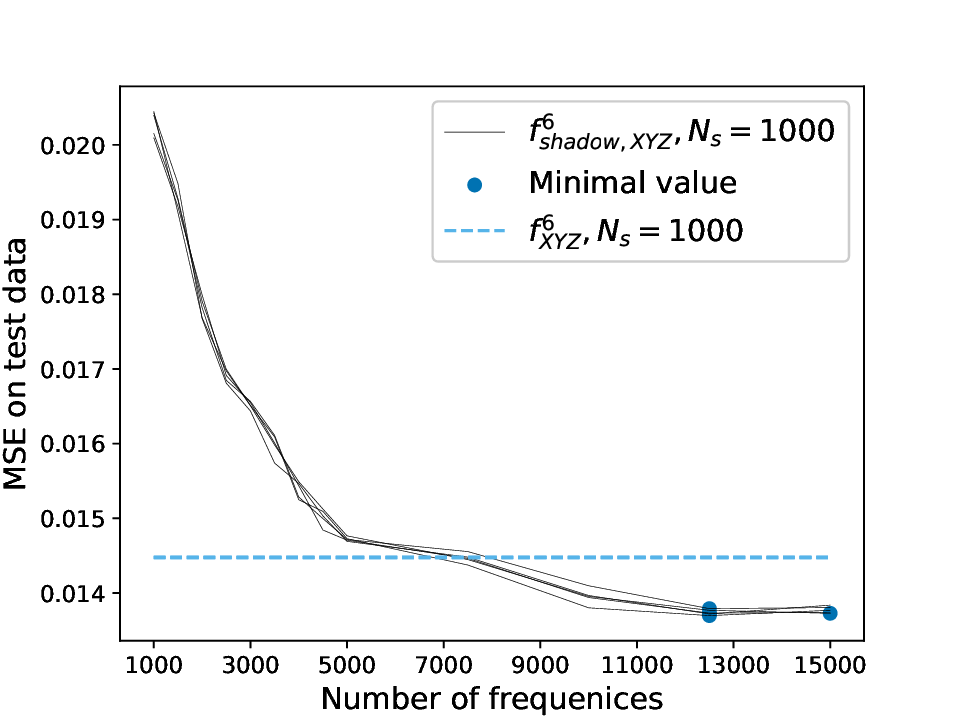} 
\includegraphics[width =  0.45\linewidth]{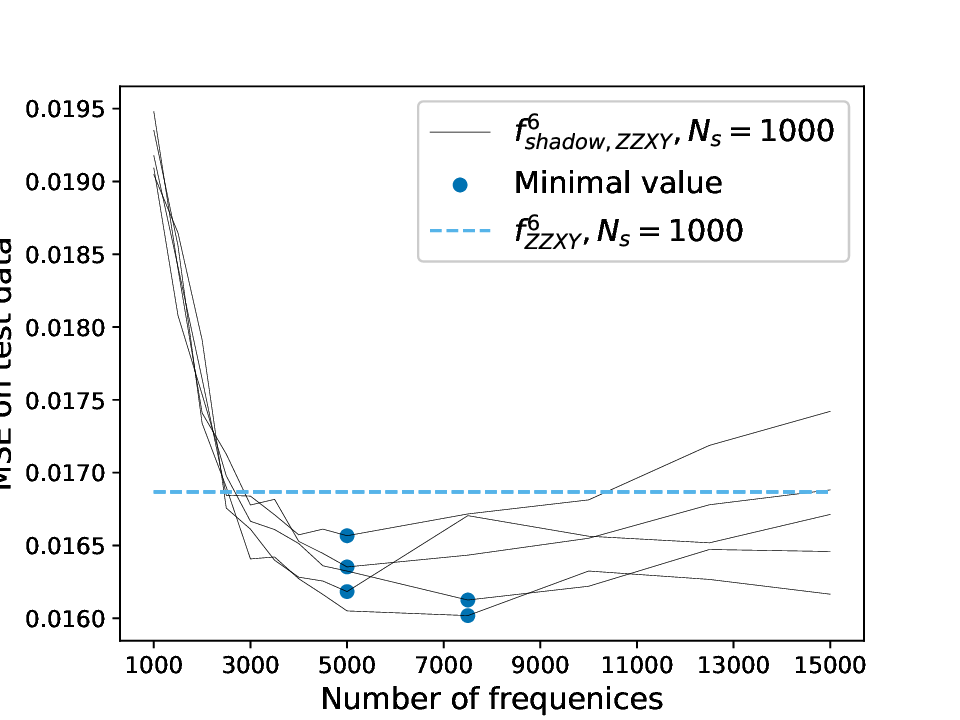} 
\caption{MSE on test data $\mathcal{D}$ of QML models compared to MSE of Fourier-based shadow models for the two architectures (left: XYZ, right: ZZXY) as functions of frequencies kept (x-axis), with  1000 shots used during training and variance regularization. Grey:  Each line corresponds to a Fourier series as a function of retained frequencies (x-axis), with a distinct set of coefficients each computed with $N_s$ number of shots to account for the variability due to shot noise. Blue dashed lines: MSE's of QML model evaluated with 1000 shots. The evaluation was repeated three times, but the MSE's are nearly indistinguishable. Blue dots represent the minimum of the MSE of different shadow models. The truncation of the frequencies spectrum yields a lower MSE than a direct evaluation of the QML model. Evaluated on test data set $\mathcal{D}$.} \label{fig:fft_1000t_1000s}
\end{center}
\end{figure}

Next,  we consider the influence of finite sampling noise on the shadowing process using the quantum Fourier model, as described in Section~\ref{sec:methods_shots}. First, we discuss the case where the model was trained analytically, i.e., with infinite shots, then a training regime with shot noise and variance regularization.
In Figure \ref{fig:fft_inft_1000s} we show the results for models $f_{ZZXY}$ and $f_{ZXY}$ trained with infinite shots but evaluated with $N_s = 100, 1000$ shots. Each line in the figures corresponds to a Fourier series with a distinct set of coefficients computed with the variability of shot noise taken into account, but shows how the truncation of this Fourier series affects the MSE. We can see that the truncation has a clear error mitigating effect compared to the QML model evaluated on the same test set. 
\\
Since some error is to be expected when training in a noiseless regime but evaluating in a noisy regime, we next repeat the shadowing process for the QML models trained with 1000 shots with added variance regularization, see Figure \ref{fig:fft_1000t_1000s}. The error mitigating effect lessens, but can still be observed. The construction of the shadow model is repeated five times, while the evaluation of the QML model is repeated three times. The MSE of the QML model shows little variability, which can most likely be attributed the size of the test set.  Meanwhile for the shadow models, the variability of the coefficients of the Fourier series and truncation can be seen well, especially when only 100 shots are used. This variability decreases when the model is trained with shot noise and variance regulation.
That the shadow models of the circuit $f_{XYZ}$ show a smaller variability than for $f_{ZZXY}$ can most likely be attributed to the spectrum of the former being significantly larger than the spectrum of the latter. Since we need to evaluate the same number of grid points to compute the coefficients as there are frequencies in the spectrum, overall more evaluations of the quantum circuit are needed for the model $f_{XYZ}$ to compute the Fourier series than for the model $f_{ZZXY}$. 

Heuristically, an explanation for this observed reduction in the MSE might be, that if the function sufficiently concentrates in the Fourier spectrum, summing over fewer noisy coefficients outweighs the error caused by the truncation, leading to an overall lower MSE, see Appendix \ref{app:noise}.  
For general applications, we cannot expect to know apriori either where the function of the trained models concentrates in the Fourier series, nor the specific variances of the finite sampling error. Therefore, a theoretical framework to find an optimal threshold to truncate the spectrum following the considerations in Appendix \ref{app:noise} is challenging. Otherwise, the threshold has to be determined experimentally as demonstrated here for the cloud cover model. This may however impact generalization skills of the models depending on the size of the data set.
Also, we evaluate the shadowing process here only by its reduction of the MSE, and leave the effects on other statistical benchmarks and on online-performance to future research.

\subsubsection{Tests on Euro-Q-Exa}  \label{sec:results:qfm:iqm}

To test the applicability of  simulator-trained quantum machine learning models on quantum hardware, we also run experiments on Euro-Q-Exa, a quantum computer located at the Leibniz Supercomputing Center based on the 53-qubit superconducting IQM Radiance technology. We do not perform any additional error correction.
We use the circuit $f^6_{ZZXY}$ with the parameters trained with finite sampling noise and  variance regularization as used in \cite[Section 5.2]{Pastori.14.02.2025}, the smaller of the circuits, with two encoding layers leading to a spectrum with $15\,625$ frequencies, such that the construction of the shadow model  on quantum hardware is feasible. While \cite{Pastori.14.02.2025} trained the circuit in an idealized setting with only finite sampling, we conduct here both the direct QML evaluation and the shadow model construction Euro-Q-Exa. 

The depth of the original circuit is 17 and compiles to the native gate set $(RX,RY,CZ)$ to 51 circuit layers. The test set data we use for a direct evaluation of the QML model is the reduced test set $\mathcal{D}_r$ due to limited quantum hardware resources.

\begin{figure}[h]
\begin{center}
\includegraphics[width = 10cm]{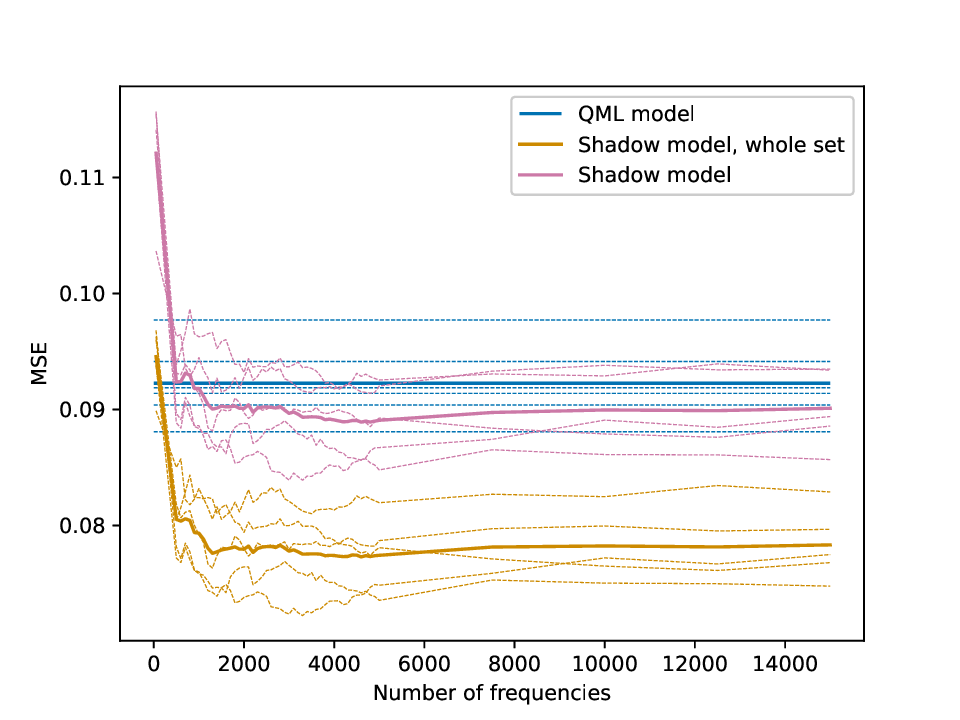}
\caption{MSE of the shadow model compared to evaluation of the QML model $f^6_{ZZXY}$ on Euro-Q-Exa. Dashed lines describe different runs of the experiment repeated over several days, i.e., on with different calibration, solid lines the average of the MSE over the corresponding experiments. Experiments on the reduced data set $\mathcal{D}_r$ are shown in blue and pink, while the MSE of the shadow model on the whole test data set $\mathcal{D}$ is plotted in yellow. On average, the shadow model has a lower MSE, however variance between different runs is still large. The effect of the frequency truncation on the MSE generalizes to the whole test data set.} \label{fig:hardware_qfm}
\end{center}
\end{figure}

We approximate each expectation value with $N_s = 1000$ shots. In Figure \ref{fig:hardware_qfm}, we compare the evaluation of the shadow model on the test data set to the evaluation of the test data directly on the hardware. Several runs were conducted over several different days to account for differences due to the calibration of the hardware. On average, the shadow model  yields a lower MSE than the direct evaluation of the QML model on the hardware. However, it is difficult to say whether this is due to the hardware calibration or an effect of the shadowing process. 
Since the test data set $\mathcal{D}_r$ is only a fraction of the whole test data set $\mathcal{D}$ the figure also shows the MSE of the shadow model on the $\mathcal{D}$. While we cannot compare it to a direct evaluation of the QML model on $\mathcal{D}$ due to limited hardware availability, Figure \ref{fig:hardware_qfm} indicates that the effect of the truncation generalizes well. Please note that the lower error on $\mathcal{D}$ is the result of the composition of the test data, and not an indication of an improved performance by the shadow model compared to the evaluation of the reduced test set. 

\begin{figure}[h]
\begin{center}
\includegraphics[width =  0.48\linewidth]{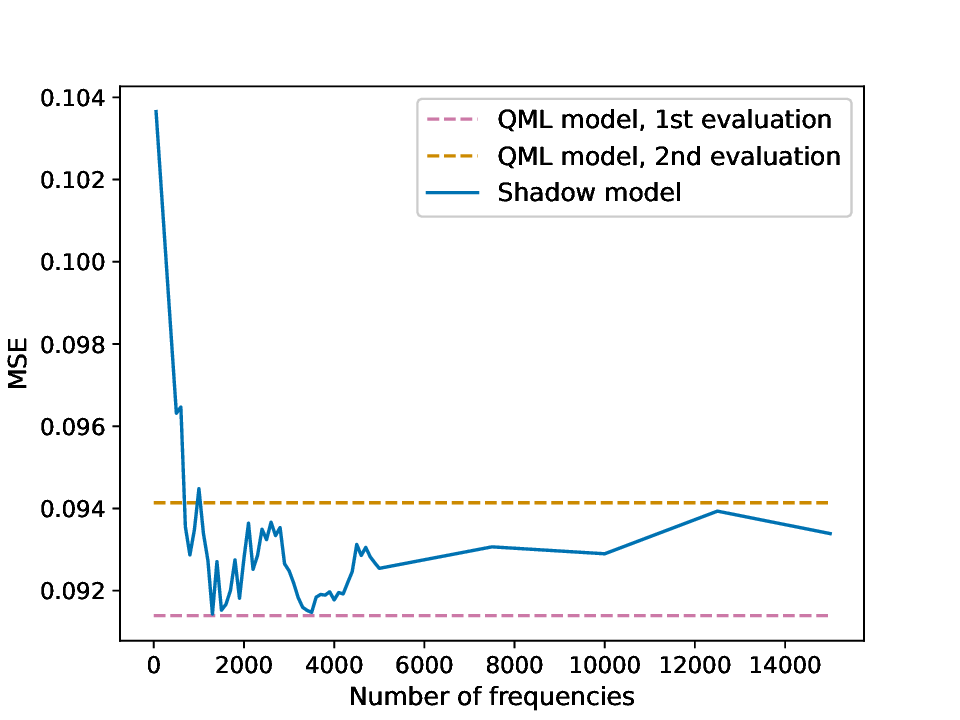}
\includegraphics[width =  0.48\linewidth]{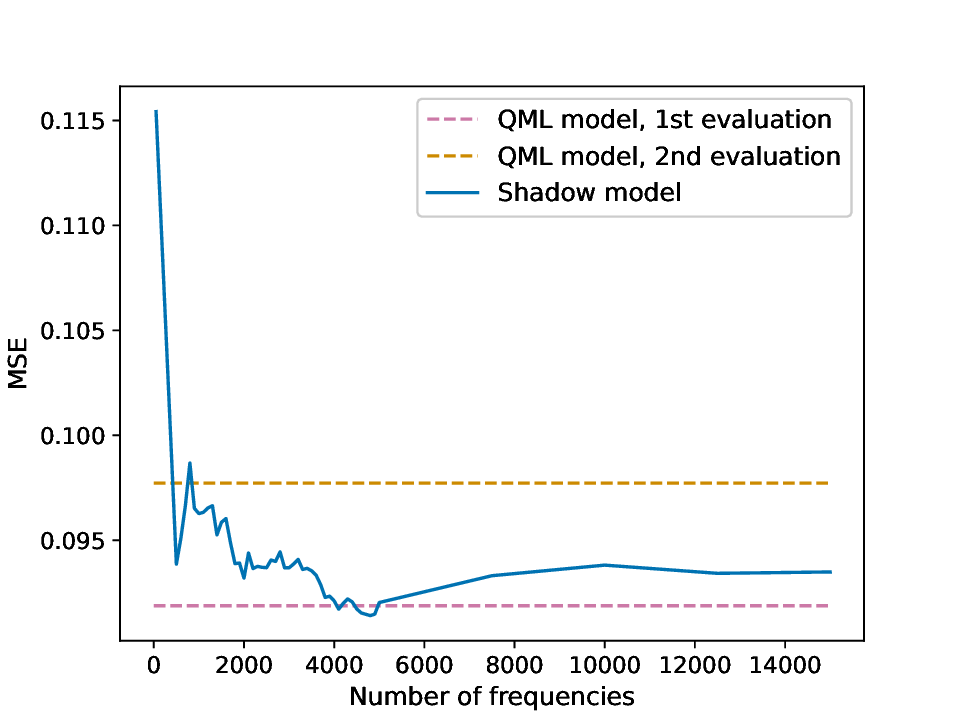}
\caption{MSE of the shadow model (solid line) compared to evaluation of QML model (dashed lines) on Euro-Q-Exa, on the test data set $\mathcal{D}_r$. For comparability, experiments where run on same days, i.e., same calibration (left: on 21st of April 2026, right: 23rd of April 2026) with the following order: First, an evaluation of the QML model (pink line), then a construction of the shadow model (blue graph), and lastly another evaluation of the QML model (yellow line). }\label{fig:hardware_qfm_sameday}
\end{center}
\end{figure}

 To further investigate the variability of the calibration a series of experiments was conducted on the same days, i.e., with the same calibration, see Figure \ref{fig:hardware_qfm_sameday}.
Within the same day, we first run an evaluation on the test data set, then construct a shadow model, and the third experiment is again an evaluation of the test data set. The truncation of the shadow model still seems to have an error mitigating effect, however much of it depends on the calibration.

\begin{figure}[h]
\begin{center}
\includegraphics[width =  0.75\linewidth]{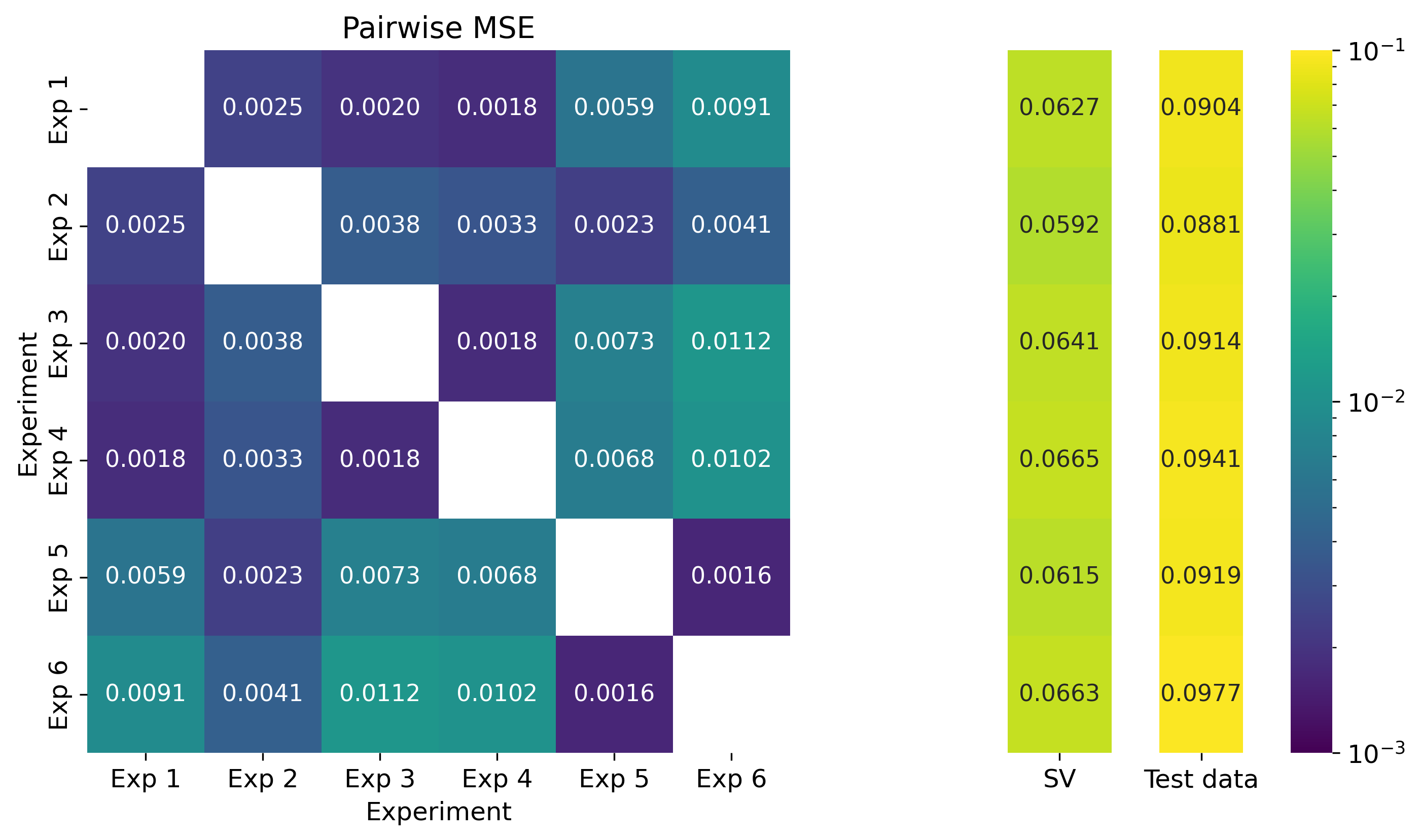}
\caption{Heat map for the pairwise MSE (left) between the predicted cloud cover on the test data set $\mathcal{D}_r$ among the runs of the QML model on Euro-Q-Exa (IQM Radiance) for experiments Exp $i, i \in \{1, \dots,  6\}$ as well as the MSE to the state vector simulation $SV$ with finite sampling noise (1000 shots) (middle) and MSE to the test data (last column). The two smallest values are the MSE between runs on the same day (Experiments (3,4) and Experiments (5,6)), but not significantly so. The discrepancy between  experiments run on hardware is an order of magnitude lower than the difference between hardware experiments and the state vector simulation with simulated finite sampling noise.}\label{fig:hardware_mse_headmap}
\end{center}
\end{figure}

In Figure \ref{fig:hardware_mse_headmap} we show the pairwise MSE between different predictions of the cloud cover by the QML model from experiments on the quantum hardware (Experiments $i$, $i \in \{1, \dots, 6\}$, as well as the MSE of the predictions on hardware to the state vector simulation with finite sampling noise and the MSE to the output of the test data set. Despite the stochasticity of the noisy quantum circuit, the error between predictions on the quantum hardware is significantly smaller than the error to the state vector simulation or the test data set, and varies similarly on days with the same calibration. This suggests that at least some of the error induced by hardware noise can in principle be regulated by training on hardware or adding a suitable regularization to a state-vector simulated training regime, similar to the variance regularization to mitigate finite sampling noise.

\section{Discussion} \label{sec:discussion}

In this work, we compared two different, constructive approaches to shadow QML models. First, a Fourier-based approach in Section \ref{sec:methods:qfm}, which uses the fast Fourier transform to compute the coefficient of the partial Fourier series, which can then be truncated to reduce the size of the shadow models. The second approach in Section \ref{sec:methods:interpolation} uses a quasi-interpolation as a circuit-agnostic alternative to the Fourier-based approach, in case the Fourier spectrum is unknown or not uniform, such that the FFT is not applicable.
We apply these methods to the QML model for cloud cover developed in \cite{Pastori.14.02.2025} and show that they are suitable to approximate the QML model in a noiseless setting. In a noisy setting the shadow models are even able to outperform the QML models and reduce the noise caused by finite sampling under suitable conditions. We evaluate one of the QML models and construct its shadow models also on the quantum computer Euro-Q-Exa, where we are also able to observe this error correcting effect, even if it is difficult to separate it from differences arising from system calibration.

These methods emphasize identifying function concentration within the QFM spectrum—or other efficient basis functions— and motivates the use of shadow models despite current hardware limitations.

Obvious caveats of the proposed methods are the poor scaling or that they only work well under restricting conditions on the encoding layers. 

Regarding our specific target QML model, we have only considered the parametrization with six input features, due to the feature analysis shown in \cite{Pastori.14.02.2025}, and since the eight dimensional models were only trained with infinite shots.

Scaling is one challenge that could be circumvented by using more computational resources. One could still apply these methods on the eight-qubit circuits, especially the quantum Fourier model for the $ZZXY$ architecture with a frequency spectrum the size of approximately $4 \cdot 10^6$.
While the shadowing process is quite costly, we note that this is largely in terms of classical resources. In terms of quantum resources the shadowing methods are cheaper than a direct coupling of the quantum model to the climate model.
The interpolation methods use the largest amount of quantum resources of the considered methods, since they require an evaluation of approx.  $ 6 \cdot  10^{10}$ points for a sufficient approximation according to the metric used in \eqref{eq:frac:metric}. 

Meanwhile a cell-based climate simulation with a horizontal grid solution of approximately 80~km (with 81920 cells per horizontal layer \cite{Giorgetta.2018}), vertical grid solution with 50-100 layers over 20 years with one time step per 120~s (i.e., around $5\cdot 10^6$ time steps) would require around $ 10^{13}$ calls to the parametrization. Ideally, the cloud cover parametrization is not always called, e.g., if a switch is included for when the cloud water and cloud ice content is too small for clouds to occur. However, in the worst case, the QML model has to be called around $10^{13}$ times for a single climate simulation, already exceeding the quantum resources needed for the generation of the investigated shadow models.

Another challenge to circumvent in the shadowing processes is that of the different encoding layers. If more sophisticated data encoding is used \cite{PhysRevA.109.042421,Mitarai.2018,Liao.2022}, the FFT can no longer be used. In this case, the FFT has to be a replaced by a non-uniform FFT, for which an inverse is more difficult to compute \cite{Wiedmann.05.11.2024}.
 In general, the Fourier-based approach is impractical for more complicated parameterizations needing circuits with high expressivity and relying on a high number of input features. 
 While different applications \cite{JonasLandman.,Hernicht.08.08.2025} attempt to deal with the problem of scalability, they did not work well for our problem. Further, they involve a regression/training, which might render the QML model redundant or eliminate any potential advantage that might have occurred during the training stage. 

 The interpolation methods are circuit agnostic and are not limited by the number of encoding gates and the size of the Fourier spectrum. However, the curse of dimensionality makes them impractical for higher dimensional problems.

While there are a variety of methods available to reduce (classical) machine learning models, e.g., surrogation \cite{Sturek.2025} or knowledge distillation \cite{Hinton.2015,Gou.2021}, they also usually have a training stage and seem more useful for large models. Although it is possible that these methods yield good results for QML models with less accessible frequency spectra, their potential and risk are difficult to evaluate without a specific application at hand.

There seems to be a lack of constructive interpolation methods for high-dimensional problems that make use of being able to choose the points adaptively.
Most likely this is due to there already existing a variety of machine learning methods available that can efficiently be used to solve large scale problem. Additionally, that outperforming the original (in our case quantum) model is seen as a risk is in general not a problem for classical models, and a more typical concern for quantum computing, where the research community desires to find advantages over classical computing.
Most likely the development of a shadow model will be problem-specific, and if an ML approach has to be used for more complex architectures, the design will likely need expert knowledge about the function space it spans and an adaptive sampling technique to generate the quantum advice to train the shadow model on efficiently. \\
However limiting in its application, the experiments for the cloud cover parametrization show the benefit of the shadowing process as an error mitigating strategy additionally to variance regularization in the training stage. An additional caveat here is that the error mitigation relies on having a large enough test data set to estimate the benefit of the truncation even if experiments indicate that the effect of the truncation generalizes well to a larger test set.
The hardware experiments show that while the hardware noise other than finite sampling noise still largely depends on the calibration,  this noise seems to be consistent enough, so that it could perhaps be suppressed by a suitable training stage on noisy intermediate-scale quantum hardware or by adding regularization to a shadowing process. It would be interesting to see how this transfers to different hardware architectures.
Also, we remark, that while the truncation has shown beneficial effects on the MSE, we have not yet investigated its effect other statistical benchmarks and the online performance.
As in \cite{Pastori.14.02.2025}, all evaluations were done offline, without coupling the shadow models to the climate model. The next step in the development of a climate model version incorporating QML-based parameterization is to see how well these parametrizations work online, and which influence the shadow models have on the stability of the climate simulations.

\ack{
This project was made possible by the DLR Quantum Computing Initiative and the Federal Ministry of Research, Space and Technology; qci.dlr.de/projects/klim-qml. V.E. was additionally supported by the
Deutsche Forschungsgemeinschaft (DFG, German Research Foundation) through the Gottfried Wilhelm Leibniz Prize awarded to Veronika Eyring (Reference No. EY 22/2-1). This work used resources of the Deutsches Klimarechenzentrum (DKRZ) granted by its Scientific Steering Committee (WLA) under project ID bd1179.
The authors gratefully acknowledge the use of the quantum system Euro-Q-Exa, co-funded by the EuroHPC JU, BMFTR (grant 13N16690), and the Bavarian State Ministry of Science and the Arts, operated by the Leibniz Supercomputing Centre (LRZ) in Garching, Germany, for providing the computational resources for this work.
Further, we gratefully acknowledge the use of Lorenzo Pastori's code for the cloud cover parametrization which this work is based on \cite{Pastori.14.02.2025}. A Large Language Model (Gemma 4) was employed to help with text-refinement and proofreading, deployed locally via Ollama.  
}

\data{
The source code supporting this research as well as the data produced from Euro-Q-Exa is hosted in the GitHub repository \url{https://github.com/EyringMLClimateGroup/keller26NewJPhys_QNN-cloudcover-shadow-models}.
The test data used is hosted at \url{https://doi.org/10.5281/zenodo.21455691}.
}

\printbibliography

\appendix

\section{Quasi-interpolation on sparse grids} \label{app:sparse_grids}

\subsection{Methods}
\begin{figure}[t]
\begin{center}
\includegraphics[width = 7cm]{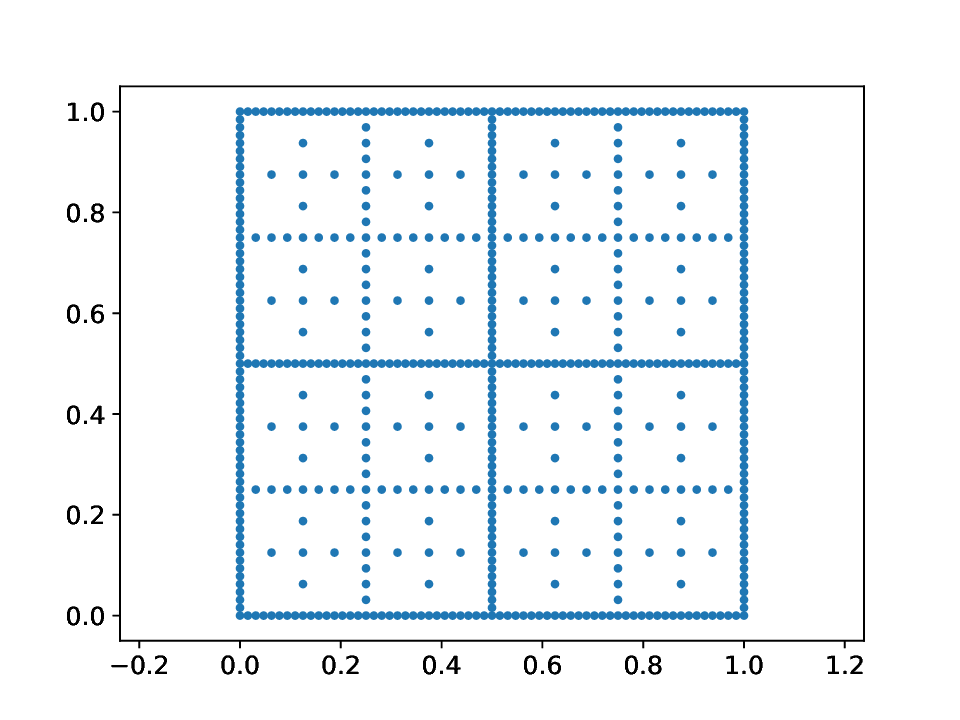}
\caption{Sparse grid $W_{(6,2)}$ on $[0,1]^2$.} \label{fig:sparse_grid2d}
\end{center}
\end{figure}

Sparse grids are defined through the layering of directionally uniform grids $W_\ell $ for a refinement parameter $ \ell \in \mathbb{N}^d$, such that the grid $W_{\ell}$ has in the direction of each  dimension $i$ an equidistant distance $h_i = 2^{-\ell_i}$, for $i \in \{1, \dots, d\}$.
The sparse grid is then defined by
\begin{equation}
W_{(n,d)} = \bigcup_{\vert \ell \vert_1 = n + d -1} W_\ell,
\end{equation}
see \cite{Gao.2024,Bungartz.2004}.
An example for a two-dimensional grid with $d = 2, n = 6$ is shown in Figure \ref{fig:sparse_grid2d}.

Interpolation on a sparse grid $W_{(n,d)}$ is usually defined by summing over suitable weighted interpolation operators $I_{\ell}$ on $W_\ell $. This yields good results in lower dimensions, however for higher dimensions, iterating over all $\ell \in \mathbb{N}^d$ s.t. $\vert \ell \vert = n + d -1$ reduces practicability. 
For a rudimentary test to compare the error to the interpolation error on the full grid, we used SciPy's \texttt{RBFInterpolator} \cite{2020SciPy-NMeth}, which is  computationally expensive (and technically not regression free in its implementation). Since this showed no advantage in terms of a convergence rate, no further research was put into this approach.

\subsection{Results}
In this section, we show the results of the shadow models using the interpolation methods using sparse grids.
We compare here the mean squared errors of the test data set for the quasi-interpolation operator \eqref{eq:qi_full}, denoted by $\mathcal{I}f$ on the full grid, and the RBF interpolation operator, see Section \ref{sec:methods:interpolation},  denoted by $\mathcal{I}_{sparse}f$ on the sparse grid, to the mean squared error of the respective 6-qubit QML models (denoted by $f^6_{XZY}$ and $f^6_{ZZXY}$).

\begin{figure}
\includegraphics[width = 0.5\linewidth]{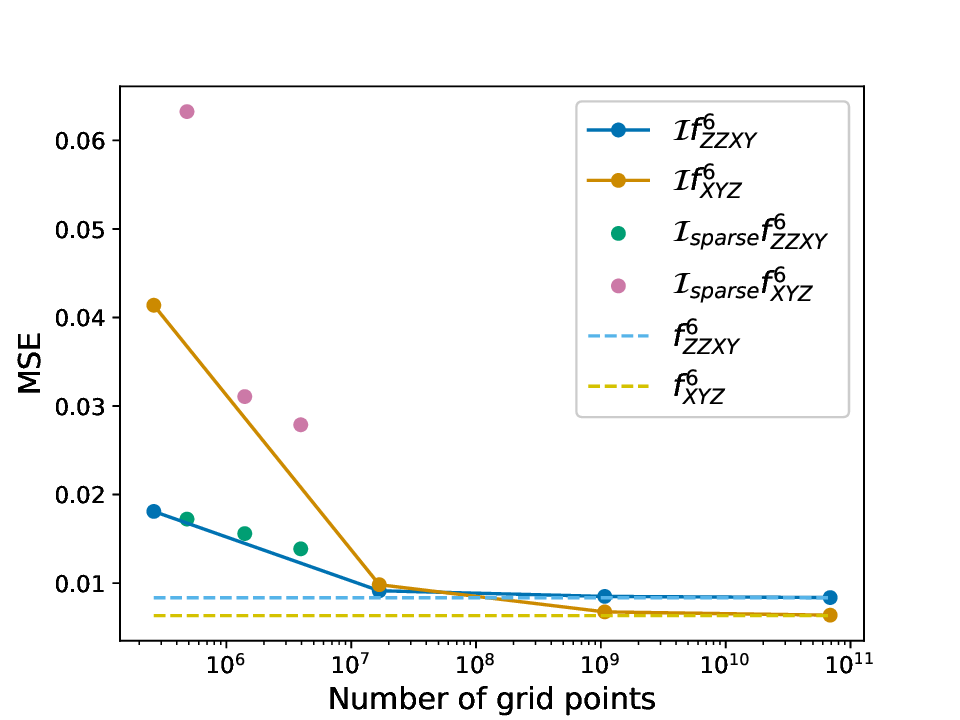}
\includegraphics[width = 0.5\linewidth]{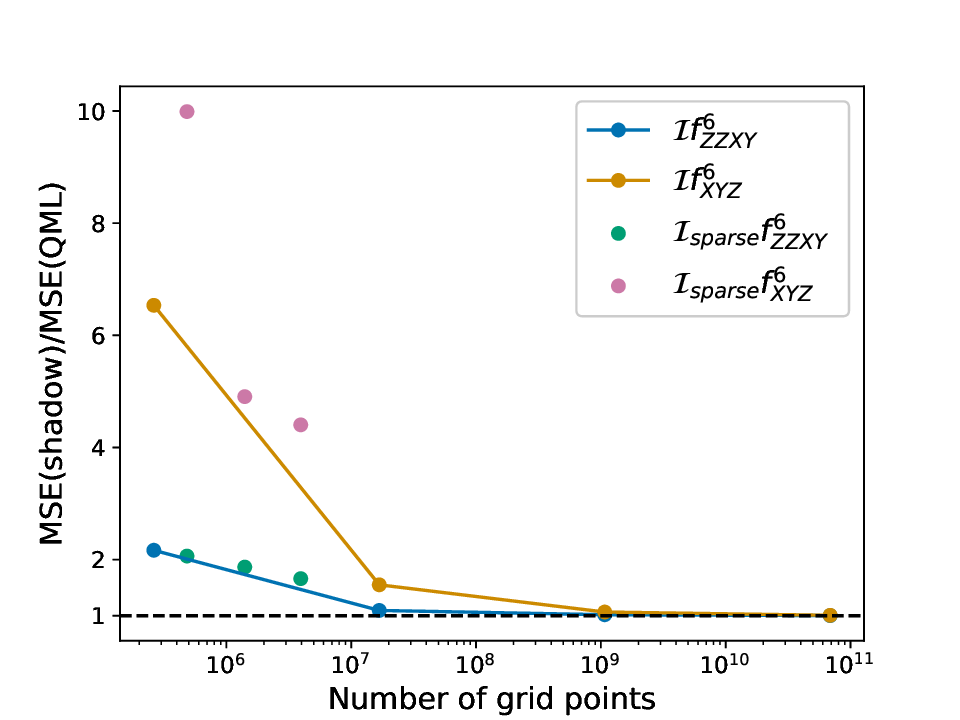}
\caption{Left: MSE of the QML models $f^6_{ZZXY},f^6_{XYZ}$ on the cirrus test data set $\mathcal{D}_c$ (dashed lines) compared to MSE of interpolants on full grids  $\mathcal{I}f^6_{ZZXY},\mathcal{I}f^6_{XYZ}$  (solid lines) and sparse grids  $\mathcal{I}_{sparse}f^6_{ZZXY},\mathcal{I}_{sparse}f^6_{XYZ}$ (no lines)  plotted against the number of points in the grid we interpolate on. Each grid refinement is indicated by a dot. Right:  Ratio of the MSE of interpolants and MSE of QML models, $\frac{MSE(\mathcal{I}f_{\cdot})}{MSE(f_{\cdot})}$ as a function of the number of grid points. The closer the fraction is to the value 1 (indicated by the dashed line as guide to the eye), the better the shadow model approximates the QML model. The sparse grid interpolants (no lines) seem to have a similar, if not worse, convergence rate than the interpolant on the full grid (solid lines) compared by the number of grid points.  QML models were trained and evaluated in a noiseless regime.}\label{fig:qi_noiseless_sparse}
\end{figure}

The interpolation on the sparse grid has a comparable, if not worse, performance as on the full grid when comparing the mean squared error (MSE), taking into account the needed grid points, see Figure \ref{fig:qi_noiseless_sparse}. Since no improvement in the rate of convergence was gained from using the sparse grid, and the implementation of sparse grid interpolators in higher dimensions is computationally expensive, this line of research was not considered further.

\section{The influence of truncation on the error caused by finite sampling noise}\label{app:noise}
In the experiments in Section \ref{sec:results} we observe that truncating the Fourier series to the largest coefficients has an error-mitigating effect if we leave the idealized setting and approximate expectation values with finite sampling noise.
To identify where the observed improvement originates from, we consider the calculations for a simplified one-dimensional problem from \cite{Lai.2026}, which incorporates an interpolation strategy for the Fourier basis into the optimization to improve variance. This can be seen as an alternative approach to the variance regularization to reduce finite sampling noise proposed in \cite{Kreplin.2024}, and highlights as well the importance of considering finite sampling noise in the optimization of parametrized quantum circuits.

When evaluating a parametrized quantum circuit $f_Q(x)$ by estimating the expectation value from $N_s$ samples, we can assume due the central limit theorem that 
\begin{equation}
f^{N_s}_Q(x)= f_Q(x) + \mathcal{N}\left(0, \frac{\sigma^2(x)}{N_s}\right).
\end{equation}
For a more detailed derivation of this, we refer to \cite[Methods]{Lai.2026}. Then, under the (in general not satisfied) assumption of constant variance $\sigma^2$, and using similar arguments for the Fourier coefficients for equidistant points, we infer that the interpolated coefficients and the QFM are distributed according to 
\begin{equation}\label{eq:variance_fft}
\begin{aligned}
\widetilde{a}_{\omega} &\sim \mathcal{N}\left(a_\omega,\frac{\sigma^2}{N_s(2L+1)}\right), \\
\widetilde{b}_{\omega} &\sim \mathcal{N}\left(b_\omega,\frac{\sigma^2}{N_s(2L+1)}\right), \text{ and} \\
\widetilde{f}^{N_s}(x) & \sim \mathcal{N}\left(f_Q(x),\frac{\sigma^2}{N_s}\right).
\end{aligned}
\end{equation}
Of particular importance are the computations preceding equation \cite[Equation 35]{Lai.2026} and Lemma 11 in the supplementary material of \cite{Lai.2026}. From this we can infer that the function truncated to the $M$ largest frequencies, with $M+1$ coefficients in total, see Section \ref{sec:methods:qfm}, has the variance
\begin{equation}
\mathrm{Var}[\widetilde{f}^{N_s}_M(x)] = \frac{(M+1)\sigma^2}{(2L+1)N_s}.
\end{equation}
So, naively speaking, we reduce the variance by summing over fewer noisy coefficients. This allows to reduce the sampling error, but adds a truncation error. If the function concentrates well enough the benefits of reducing the sampling variance can outweigh the truncation error. For a more detailed computation  we refer to \cite{Lai.2026}. For details on avoiding the assumption of constant variance see \cite[Supplementary Note 8]{Lai.2026}. In this case, the variance can also be bound, but problem-specific on the minimal and maximal variance of the sampling error at the grid points.
A similar result as \eqref{eq:variance_fft} can be extended to error mitigation effect of the interpolation strategy of the form \eqref{eq:interpolation} observed in  Section \ref{sec:results}, however a precise estimate of the variance depends on the respective basis functions.

In \cite{nair2026localtensortrainsurrogatesquantum} a truncated Taylor model is used to construct local-train surrogates, for which also an improvement in the test error is observed. It might be worth investigating if the effect of a Taylor truncation can be compared to the truncation of the Fourier series. 
Another publication that suggests truncation as an error-mitigating strategy, albeit for hardware noise, is \cite{Fontana.17.06.2022}. They suggest that it is only useful for a heuristic setting, since it assumes a sparse signal and the goal of the truncation is to filter out the relevant frequencies of the sparse signal. We show that this can extend to applications for which the function is sufficiently concentrated, if not sparse.

Another investigation of the influence of noise on the Fourier spectrum is considered in \cite{lu2026unifiedfrequencyprinciplequantum}. In particular they research the spectral bias of QNN's on lower frequencies, also in noisy settings and note the importance of spectral filtering.

\end{document}